\pdfoutput=1

\documentclass[preprint,aps,nofootinbib]{revtex4}
\usepackage{bm}
\usepackage{graphicx}

\usepackage[hyperfootnotes=false]{hyperref}
\usepackage{url}

\def\lsim{\mathrel{\raise.3ex\hbox{$<$\kern-.75em\lower1ex\hbox{$\sim$}}}}
\def\gsim{\mathrel{\raise.3ex\hbox{$>$\kern-.75em\lower1ex\hbox{$\sim$}}}}

\newcommand{\sigmav}{\ensuremath{\langle\sigma v\rangle}}

\newcommand{\bea}{\begin{eqnarray}}
\newcommand{\eea}{\end{eqnarray}}

\newcommand{\be}{\begin{equation}}
\newcommand{\ee}{\end{equation}}

\graphicspath{{figures/}}

\usepackage[usenames]{color}
\definecolor{DarkRed}{rgb}{0.5,0.0,0.0}
\definecolor{DarkGreen}{rgb}{0.0,0.5,0.0}
\definecolor{DarkBlue}{rgb}{0.0,0.0,0.5}
\definecolor{DarkMagenta}{rgb}{0.5,0.0,0.5}
\definecolor{DarkOrange}{rgb}{0.8,0.3,0.0}
\definecolor{DarkCyan}{cmyk}{1.0,0.0,0.0,0.5}

\begin{document}
\begin{flushright}
\vspace*{-0.6cm}
\vspace{-0.9cm} 
MCTP-15-17, WSU-HEP-1504, NORDITA-2015-160\\
\end{flushright}

\vspace*{-0.8cm}

\title{MSSM $\mathbf{A}$-funnel and the Galactic Center Excess:\\ Prospects for the LHC and Direct Detection Experiments\\}



\author{Katherine Freese$^{1,2,3}$}
\email{ktfreese@umich.edu}

\author{Alejandro L\'{o}pez$^3$}
\email{aolopez@umich.edu}

\author{Nausheen R. Shah$^{3,4}$}
\email{naushah@umich.edu}

\author{Bibhushan Shakya$^3$}
\email{bshakya@umich.edu}

\affiliation{\vskip0.1cm$^1$Nordita (Nordic Institute for Theoretical Physics), KTH Royal Institute of Technology and Stockholm University, Roslagstullsbacken 23, SE-106 91 Stockholm, Sweden.\\
$^2$The Oskar Klein Center for Cosmoparticle Physics, AlbaNova University Center, University of Stockholm, 10691 Stockholm, Sweden.\\
$^3$Michigan Center for Theoretical Physics, Department of Physics, University of Michigan, Ann Arbor, MI 48109, USA.\\
$^4$Department of Physics and Astronomy, Wayne State University, Detroit, Michigan 48201, USA}


\begin{abstract}

The pseudoscalar resonance or ``$A$-funnel" in the Minimal Supersymmetric Standard Model~(MSSM) is a widely studied framework for explaining dark matter that can yield interesting indirect detection and collider signals. The well-known Galactic Center excess (GCE) at GeV energies in the gamma ray spectrum, consistent with annihilation of a $\lsim 40$ GeV dark matter particle, has more recently been shown to be compatible with significantly heavier masses following reanalysis of the background. In this paper, we explore the LHC and direct detection implications of interpreting the GCE in this extended mass window within the MSSM $A$-funnel framework.  We find that compatibility with relic density, signal strength, collider constraints, and Higgs data can be simultaneously achieved with appropriate parameter choices. The compatible regions give very sharp predictions of $200-600$ GeV CP-odd/even Higgs bosons at low $\tan\beta$ at the LHC and spin-independent cross sections $\approx 10^{-11}$ pb at direct detection experiments. Regardless of consistency with the GCE, this study serves as a useful template of the strong correlations between indirect, direct, and LHC signatures of the MSSM $A$-funnel region.

\end{abstract}


\maketitle

\section{Introduction and Motivation}\label{intro}

The Galactic  Center (GC) of the Milky Way galaxy is the densest dark matter region in our vicinity and has long been earmarked as the most promising target for searches of dark matter (DM) signals. Intriguingly, recent years have seen a persistent and statistically significant excess in the gamma ray spectrum peaking at $2-5$ GeV originating from the GC, above what is predicted from known sources and conventional astrophysics  \cite{Goodenough:2009gk,Hooper:2010mq,Boyarsky:2010dr,Hooper:2011ti,Linden:2012iv,Abazajian:2012pn,Hooper:2013rwa,Gordon:2013vta,Abazajian:2014fta,Daylan:2014rsa,Zhou:2014lva,Calore:2014xka}. The signal was initially reported to be compatible with $\sim 40 \,(10)$ GeV dark matter annihilating into $b\bar{b}$ ($\tau\tau$), with an annihilation cross section $\langle \sigma v \rangle \sim \mathcal{O}(10^{-26})$ cm$^3$/s. Since this is approximately the annihilation cross section expected of a thermal relic, a dark matter interpretation of this excess presents itself as a very tantalizing possibility. This prospect has been explored by many authors in various contexts (see, for instance Refs.~ \cite{Daylan:2014rsa,Abazajian:2014fta,Calore:2014nla,Abdullah:2014lla} and references therein), including the Minimal Supersymmetric Standard Model~(MSSM) \cite{Agrawal:2014oha, Cahill-Rowley:2014ora,Caron:2015wda,Bertone:2015tza}. More recently, it has  been shown that this excess might be attributable to unresolved point sources \cite{Lee:2015fea,Bartels:2015aea,Gaggero:2015nsa}, although a conclusive verdict has not been reached. 

Recently, the Fermi-LAT Collaboration has presented an analysis of the region around the GC with four different variants of foreground/background models, finding, for every variant, significant improvements in the agreement with data when an additional component centered at the GC with a peaked profile~(NFW, NFW-contracted), {\textit{i.e.}} a dark matter-like spectrum, was included in the fits \cite{simonatalk, TheFermi-LAT:2015kwa} (see also Ref.~\cite{Calore:2014xka} for an attempt at accounting for systematic uncertainties in the background). From a dark matter perspective, a recent study \cite{Agrawal:2014oha} found these additional components for the four choices of background models to be compatible with several annihilation channels $(WW,ZZ,hh,t\bar{t})$ and significantly higher DM masses ($165$ GeV for $b\bar{b}$, $310$ GeV for $t\bar{t}$) than previously thought possible. Similar conclusions were also reached in Refs.~\cite{Caron:2015wda} and \cite{Bertone:2015tza}, which reported that a higher mass ($175-200$ GeV) dark matter annihilating into $t\bar{t}$ could give reasonable fits to the signal.

This relaxation of the allowed range of dark matter masses compatible with the GC excess~(GCE) has particularly interesting implications for MSSM dark matter, as it opens up the possibility of explaining the signal with the well-known pseudoscalar resonance or ``$A$-funnel" mechanism, where the dark matter relic density is set by resonant $s$-channel annihilation through the pseudoscalar $A$, with $m_A\approx2m_\chi$ (\,$\chi$ represents the lightest neutralino, which is the dark matter candidate). The pseudoscalar resonance has been studied in connection with the GCE outside the MSSM in Refs.~\cite{Berlin:2015wwa, Cheung:2014lqa, Ipek:2014gua}; however, realizing the mechanism in the MSSM is of particular interest given that the MSSM remains one of  the most familiar and widely studied Beyond the Standard Model (BSM) theories. Previous fits to the GCE with $m_\chi\lsim 50$ GeV did not allow for this possibility in the MSSM due to constraints on $m_A$ from direct LHC searches \cite{TheATLAScollaboration:2013wia, Chatrchyan:2012vca} (although this constraint can be circumvented in the the Next-to-Minimal Supersymmetric Model (NMSSM), allowing for an NMSSM explanation of the GCE \cite{Cheung:2014lqa, Berlin:2014pya}). This incompatibility is lifted if, as discussed in Ref.~\cite{Agrawal:2014oha}, $m_\chi\lsim165 ~(310)$ GeV annihilates into $b\bar{b}~(t\bar{t})$, allowing for $m_A$ large enough to evade collider constraints. 

The aim of this paper is to explore whether, given this wider range of allowed masses, the MSSM pseudoscalar resonance can give reasonable fits to the GCE, consistent with stringent constraints from relic density, indirect/direct detection, collider search limits, and Higgs data. Since the mechanism requires a light ($\sim 200-500$ GeV) pseudoscalar, the SM-like nature of the 125 GeV Higgs boson is particularly constraining as the heavier CP-even Higgs is at the same mass as the pseudoscalar and can mix with the 125 GeV Higgs, resulting in deviations from SM-like properties inconsistent with measurements. For such light, non-decoupled heavier Higgs bosons, the Higgs sector needs to be ``aligned" \cite{Carena:2013ooa,Carena:2014nza,Gunion:2002zf,Asner:2013psa,Haber:2013mia,Craig:2013hca} to maintain SM-like properties for the 125 GeV mass eigenstate. As we will show in this paper, this can indeed be achieved while simultaneously satisfying all other DM requirements.

A successful realization of neutralino dark matter along with the GCE through the pseudoscalar resonance requires very precise choices of parameters in order to simultaneously achieve resonant annihilation, the Higgs mass, and alignment in the Higgs sector (this is also the reason why extensive scans in the MSSM parameter space \cite{Agrawal:2014oha, Cahill-Rowley:2014ora,Caron:2015wda,Bertone:2015tza} fail to uncover it as a viable explanation of the GCE). It is nevertheless worthwhile to pursue this direction for several reasons. First, the $A$-funnel is one of several ``traditional" mechanisms in the MSSM that have been widely studied for a long time, and its compatibility with a possible DM signal is therefore of considerable interest. Second, while most scenarios put forward to explain the GCE could potentially be constrained by stringent spin-independent direct detection limits (indeed, avoiding these limits itself involves some nontrivial fine-tuning of parameters in supersymmetric models \cite{Perelstein:2011tg,Amsel:2011at,Perelstein:2012qg}), the $A$-funnel naturally gives small direct detection cross sections and is automatically safe from these bounds. Most importantly, the framework is eminently predictive, giving very specific predictions for heavy Higgs bosons that will be probed at the 13 TeV LHC and future colliders, as well as direct detection cross sections that may be probed by the next generation of experiments. Independent of these considerations, and independent of the applicability to the GCE, this study serves as a valuable template of the conditions necessary for the existence of a light pseudoscalar in the MSSM together with indirect detection signals of dark matter via the A-funnel.  

The outline of the paper is as follows. Section II introduces the parameter space relevant for the study and discusses dark matter aspects such as the annihilation cross section and relic density. Section III is devoted to a discussion of various constraints from direct detection, indirect detection, collider constraints, Higgs data, and vacuum metastability. Section IV presents the details of our scans and the best fit regions to the GCE. Predictions for the 13 TeV LHC and future direct detection searches are presented in Section V. We summarize our results in Section VI. The Appendices contains additional details on the MSSM parameters and fits to the GCE.

\section{The MSSM Pseudoscalar Resonance: Dark Matter Aspects}

In $R$-parity conserving supersymmetric models, the lightest supersymmetric particle~(LSP) is stable. If it is also neutral, it can be a dark matter candidate. In the MSSM, the LSP is often assumed to be the lightest of the neutralinos, the neutral superpartners of the gauge bosons and Higgs bosons~(Bino, Wino and Higgsinos respectively). The Wino and the Higgsinos tend to annihilate too efficiently to explain the observed dark matter abundance. However, the Bino can yield the correct relic density via various mechanisms, including resonant annihilation via the pseudoscalar, and has long been regarded as the favored dark matter candidate.   
  

We perform our study in the phenomenological MSSM (pMSSM) \cite{Djouadi:1998di}, which is defined in terms of 19 parameters, which are taken to be independent at the weak scale. Of these, our analysis will be entirely determined by the following seven parameters:

\begin{itemize}
\item $M_1$, the Bino mass parameter. The dark matter is mostly Bino, so this is also approximately  the mass of the dark matter candidate $m_{\chi}\approx M_1$.
\item $\mu$ parameter. This is the Higgsino mass, and controls the Higgsino fraction in the dark matter particle $\chi$. As we will see later, the relic density, signal strength, and direct detection cross section all depend sensitively on this fraction.  
\item $\tan\beta$, the ratio of the up- and down-type Higgs vacuum expectation values (vevs).
\item $m_A$, the heavy Higgs mass. This is the mass of the pseudoscalar that mediates the resonance (hence $m_A\approx 2 m_\chi$) as well as the mass of the heavier scalar, which feeds into Higgs phenomenology and expected direct detection cross-sections. 
\item $m_{Q_3},m_{u_3}$, the left and right handed stop masses, which contribute significantly to  the mass of the observed 125 GeV Higgs boson. In this paper we take the stop mass scale $M_S^2\equiv m_{Q3}^2 =  m_{u3}^2$.
\item $A_t$, stop trilinear coupling. This determines the mixing in the stop sector and is again a relevant parameter for the mass of the observed Higgs boson.
\end{itemize}
All other masses, such as the other gaugino (wino and gluino) and sfermion masses, are assumed to be heavy and decoupled from the analysis. 

\subsection{Dark Matter Composition}
The lightest neutralino in the MSSM is a combination of the Bino, Wino, and neutral Higgsinos:
\be
\chi=N_{11} \tilde{B}+N_{12} \tilde{W}+N_{13} \tilde{H_d}+N_{14} \tilde{H_u}\; .
\ee
As mentioned above, we are mainly interested in the region of parameters where the lightest neutralino is predominantly a Bino, hence $N_{11}\sim 1,\, N_{12}=0,\,$ and $ N_{13},N_{14}\ll 1$. In this regime, the Bino mass parameter $M_1$ and  the neutralino components are approximately\,\cite{Cheung:2014lqa}
\begin{eqnarray}\label{N_components}
M_1&=&m_{\chi }+\frac{m_Z^2 s_W^2 \left(\mu  s_{2\beta}+m_{\chi }\right)}{\mu^2-m_{\chi }^2},\nonumber\\
\frac{N_{13}}{N_{11}}&=&\frac{m_Z s_W s_\beta }{\mu ^2-m_{\chi }^2}  \left(\mu  +\frac{m_{\chi }}{t_\beta}\right) \sim \frac{m_Z s_W  }{\mu} s_{\beta},\nonumber\\
\frac{N_{14}}{N_{11}}&=&-\frac{m_Z s_W c_\beta }{\mu^2-m_{\chi }^2} \left(\mu +t_\beta m_{\chi }\right) \sim -\frac{m_Z s_W}{\mu} c_{\beta}\left(1+t_{\beta}\frac{m_\chi}{\mu}\right),\nonumber\\
N_{11}&=&\left(1+\frac{N^2_{13}}{N^2_{11}}+\frac{N^2_{14}}{N^2_{11}}\right)^{-1/2}.
\end{eqnarray}
Here, $s_\theta, c_\theta$ denote $\sin\theta, \cos\theta$ respectively and $m_\chi$ is the dark matter mass.

\subsection{Relic Density and Signal Strength}

Both the relic density and the present day annihilation cross section are driven by the process $\chi\chi\rightarrow f\bar{f}$ with the pseudoscalar $A$ in the $s$-channel (we are interested in the case where the fermion $f$ is either $b$ or $t$ for compatibility with the GCE). When the process occurs close to resonance, it is well-known that the annihilation cross-section in the early universe (which sets the relic density at the time of freeze-out) is substantially different from that at present times (which sets the signal strength fitting the GCE) due to thermal broadening of the resonance during the former stage~\cite{Griest:1990kh}. Thus, with appropriate parameter choices, one can scale the relic density and the present annihilation cross section independent of each other, thereby achieving better agreement with both measurements; this degree of freedom is not afforded in non-resonant scenarios, where these two quantities are strictly related to each other. 

To understand this interplay, consider a simplified model describing a Majorana DM particle $\chi$ coupled to a pseudoscalar $A$ through the interaction Lagrangian
\begin{eqnarray}
-{\cal L}_{\rm int} &=& i y_{a \chi\chi} A \bar \chi \gamma^5 \chi +i
y_{a ff} A \bar f \gamma^5 f .
\label{eq:simpL}
\end{eqnarray}
The entire parameter space of the model is then determined by $m_A, m_\chi, y_{a \chi\chi}$ and $y_{a f f}$. 
A crucial parameter in our analysis is the degeneracy parameter
\begin{eqnarray}\label{eq:delta}
\delta &=& |1- 4 m_\chi^2 / m_A^2|,
\end{eqnarray} 
which characterizes the proximity to the resonant regime. We are interested in scenarios where $\delta\approx 0$. 

The resonant annihilation cross-section at a given temperature $T$ is \cite{Griest:1990kh} \begin{eqnarray}
\langle \sigma v \rangle 
&\simeq & \frac{3 e^{-x \delta}    x^{3/2}\delta^{1/2}  y_{a \chi\chi}^2 y_{a ff}^2 m_\chi^2 }{\sqrt{\pi} m_{A}^3 \Gamma_A }\; ,
\label{eq:sigma}
\end{eqnarray}
where $x = m_{\chi}/T$ and $\Gamma_A$ is the decay width of $A$,
\begin{eqnarray}
\Gamma_A &\simeq & \frac{ m_A}{16\pi} (y_{a \chi\chi}^2 +6y_{a f f}^2).
\label{eq:Gamma}
\end{eqnarray}
This gives the relic abundance
\begin{equation}
\Omega h^2 = \frac{3.12 \times 10^{-12}  m_A^3 \Gamma_a}{ (\textrm{GeV})^2m_{\chi}^2 y_{a \chi\chi}^2 y_{a ff}^2 \mbox{Erfc}\left[\sqrt{x_f \delta}\right]},
\label{eq:Omega}
\end{equation} 
where $x_f$ is the value of $x$ at freeze-out. This expression can be rewritten in a more illuminating form as \cite{Cheung:2014lqa}
\begin{eqnarray}
\Omega h^2 &\sim&0.12 \left(\frac{m_A^2}{4
  m_\chi^2}\right)\left(\frac{m_A}{220\text{ GeV}}\right)^2\left[\frac{
    y_{a\chi\chi}^{-2}+(\delta/ 6) \,y_{aff}^{-2}}{10^{5}}\right] \left(\frac{\textrm{Erfc}[1.325]}{\textrm{Erfc}\left[\sqrt{x_f}\,\delta\right]}\right).
    \label{eq:Omega2}
\end{eqnarray}

Likewise, the DM annihilation cross-section today is
\begin{eqnarray}
\sigma v \big|_{v=0} &\simeq& \frac{3}{2\pi} \frac{y_{a \chi\chi}^2 y_{a f f}^2 m_\chi^2 }{(m_A^2 - 4m_\chi^2)^2 + m_A^2 \Gamma_A^2 }\, .
\label{eq:simpsigma}
\end{eqnarray}
Assuming that $m_A\sim 2m_{\chi}$ so that the second term dominates in the denominator, one obtains (for $2 m_\chi < m_a$) \cite{Cheung:2014lqa}
\begin{eqnarray}
\sigma v \big|_{v=0} &\sim& 2\times10^{-26}\text{cm}^3 \left(\frac{4
  m_{\chi }^2}{m_A^2}\right)\left(\frac{220 \text{
    GeV}}{m_A}\right)^2\frac{10^{-5}} {( \frac{y_{a\chi \chi } }{
    y_{aff}}\frac{\delta}{6}+\frac{y_{aff}}{y_{a\chi \chi
}})^2}.
\label{eq:crosssection}
\end{eqnarray}

Comparing Eq.\,\ref{eq:Omega2} and Eq.\,\ref{eq:crosssection}, it is clear that the relic density and the current annihilation cross-section can be independently scaled with judicious choices of $y_{a f f}$ and $y_{a \chi\chi} \sqrt{\delta / 6}$. In terms of the fundamental MSSM parameters, these couplings are given by:
\begin{eqnarray}
y_{a b b}=\frac{i m_b \tan{\beta}}{\sqrt{2} v},~~y_{a t t}=\frac{i m_t }{\sqrt{2} v \tan{\beta}},~~~~~~~~~~~~\\
y_{a \chi\chi}=i g_1 N_{11} (N_{14}\cos{\beta}-N_{13}\sin{\beta}), \label{achichi}
\end{eqnarray}
where $v=174$\,GeV and $g_1$ is the SM $U(1)_Y$ gauge coupling. Note from the above that a non-vanishing $y_{a \chi\chi}$ coupling requires a non-vanishing Higgsino component in $\chi$.  From the expressions for $N_{11}, N_{13}, N_{14}$ listed previously, we thus see that, for given values of $m_A$ and $\tan\beta$, the desired relic density and an annihilation cross-section consistent with the GCE can be obtained simultaneously by appropriately choosing $\mu$ and $\delta$ (equivalently, $m_\chi$).

\section{Constraints}

As mentioned in Sec.~\ref{intro}, the relevant $A$-funnel parameter space is constrained from several directions. Higgs phenomenology in our set-up is very directly linked to the GCE, hence LHC direct searches as well as the properties of the observed 125 GeV Higgs put stringent constraints on this scenario. Consistency with all collider observables can then create tension with constraints from requiring the stability of the electroweak vacuum. In addition, since the CP-even heavy Higgs $H$ is expected to be approximately degenerate in mass with $A$, contributions to the spin-independent direct detection cross-section from $H$-exchange might be relevant. Finally, there are also several current and future indirect detection experiments that can probe the process of interest in this paper. In this section we detail the current status and future prospects in all of these different directions.

\subsection {Collider and Higgs Sector Constraints}\label{HiggsIntro}

In the absence of CP-violation (which we assume in this paper), the physical spectrum of the Higgs sector  consists of two CP-even Higgs bosons, $h$ and $H$, one CP-odd state $A$, and a pair of charged Higgs bosons, $H^\pm$. Direct searches for these heavier Higgs bosons at the LHC rule out a significant part of parameter space. ATLAS and CMS direct searches for charged Higgs bosons \cite{TheATLAScollaboration:2013wia, Chatrchyan:2012vca} rule out $m_{H+}\leq 160$ GeV (recall that $m_{H+}^2=m_A^2+m_W^2$ at tree level). Likewise, there exist strong limits from searches for $A/H\rightarrow\tau\tau$\,\cite{Khachatryan:2014wca}, which provide the strongest limits, although these depend on $\tan\beta$ and can be evaded for small values of $\tan\beta$.\footnote{Light $m_A/m_H$ and heavily mixed stops (as usually needed for a 125 GeV Higgs in the MSSM) can also give large contributions to various flavor observables, for example $B_s\to \mu^+ \mu^-$ and $B\to X_s \gamma$. However, in this work we will mainly be interested in moderate to small value of $\tan\beta$, hence there is no large enhancement of these effects. Moreover, the size of these contributions are heavily dependent on the signs of various contributions (see e.g. Ref.~\cite{Altmannshofer:2012ks}), and consistency with all measured values could be obtained by tuning such cancellations.}

Beyond these direct constraints, a small $m_A$ is still in tension with Higgs data, as a light CP-even Higgs ($m_H\approx m_A$ in the MSSM) tends to mix with the 125 GeV state and cause deviations from SM-like properties. This is a particularly strong constraint in our framework and dictates what values our parameters can take, hence we will now study this constraint in some detail. 

The MSSM Higgs sector consists of two doublets, $H_u$ and $H_d$; the former couples to all the up-type fermions and the latter to the down-type fermions and charged leptons. The neutral components acquire vacuum expectation values $v_u$ and $v_d$ with $\tan\beta=t_\beta=v_u/v_d$ and $v^2=\sqrt{v_u^2+v_d^2}=174$ GeV. One can define a ``Higgs-basis", where a single field acquires all the vev:
\bea
H_{SM}&=& s_\beta H_u +c_\beta H_d, \\
H_{NSM}&=& -c_\beta H_u +s_\beta H_d,
\eea
where $s_\beta \equiv \sin\beta$,  $c_\beta \equiv \cos\beta$, $\langle H_{SM} \rangle = v$, and $\langle H_{NSM} \rangle = 0$. The couplings of these states to the SM fields are:
\bea \label{g_basis}
g^{dd/uu/VV}_{H_{SM} } &=& g^{SM} \nonumber\\
 g^{dd}_{H_{NSM}} = g^{SM} t_\beta, \qquad g^{uu}_{H_{NSM}} &=& -g^{SM}/ t_\beta \qquad g^{VV}_{H_{NSM}} = 0\; ,
\eea
where $VV, uu, dd$ refer to all vector, up-type and down-type states respectively, and $g^{SM}$ refers to the SM value of these couplings. Note that there is no coupling between the $H^0_{NSM}/A$ states and the $H^0_{SM}$ or between the gauge bosons and  $H^0_{NSM}$.

The mass eigenstates, $h$ and $H$, can be written as mixtures of the Higgs basis fields,
\bea\label{hinH}
h&=& \kappa^h_{SM} H_{SM}+\kappa^h_{NSM}H_{NSM}\;,\nonumber\\
H&=& \kappa^H_{SM} H_{SM}+\kappa^H_{NSM}H_{NSM}\;,
\eea
where $\kappa^h_{NSM}= -  \kappa^H_{SM} =c_{\alpha-\beta}\equiv \cos(\alpha-\beta)$ and $\kappa^h_{SM}=\kappa^H_{NSM}=s_{\alpha-\beta}\equiv \sin(\alpha-\beta)$, and $\alpha$ is the angle of rotation from the $(H_u,\,H_d)$ basis to the mass eigenstates.   
We want to identify the lightest CP-even mass eigenstate, $h$, with the recently observed 125 GeV scalar; given that all measurements suggest that its properties are SM-like, we also want to identify it as the SM-like field in the Higgs basis. That is, we require
\be
h_{125} =  h \approx H_{SM}.
\ee
This requirement of vanishing mixing between the $H_{NSM}$ state and the 125 GeV Higgs, corresponding to $\kappa^h_{NSM}\approx 0$, can be rewritten in terms of the fundamental parameters as ~\cite{Carena:2013ooa, Carena:2014nza}
\be \label{tbcbma}
t_\beta\; c_{\beta-\alpha}\simeq \frac{-1}{m_H^2-m_h^2}\left[m_h^2+m_Z^2+
\frac{3m_t^4 X_t(Y_t-X_t)}{4\pi^2 v^2
  M_S^2}\left(1-\frac{X_t^2}{6M_S^2}\right)\right]\, \simeq 0\; ,
\ee
where $M_S$ is the geometric mean of the stop masses and 
\begin{equation} \label{XY}
X_t\equiv A_t-\mu /t_\beta\,,\qquad\quad Y_t\equiv A_t+\mu\, t_\beta\,.
\end{equation}
Note that when the second Higgs becomes heavy ($m_H >>  m_h$), this relation is automatically satisfied; this is the familiar decoupling effect.  Otherwise, one requires alignment without decoupling \cite{Carena:2013ooa, Carena:2014nza}, brought about by an accidental cancellation in the fundamental parameters of the theory so as to satisfy Eq.\,\ref{tbcbma}. For small $t_\beta$ and $M_S\sim \mathcal{O}(1)$ TeV, large values of $ A_t/M_S$ are required to obtained an experimentally consistent Higgs mass whereas large values of $(\mu A_t )/M_S^2$ lead to close to alignment conditions \cite{Carena:2013ooa, Carena:2014nza}.

The CMS and ATLAS collaborations present both the precision measurements of the 125 GeV Higgs and the searches for $H\to WW/ZZ$ as ratios to the expectations from a SM Higgs of the same mass. The predicted rate at the LHC for the decay of the mass eigenstate $i=\{h, H\}$ into some final state $XX$ as a ratio to the SM value is given by
\be \label{RWW}
\mathcal{R}^i_{XX} =(\sigma^i/SM) \times (BR^i_{XX}/SM)\; .
\ee 
where $SM$ in the denominators denote the corresponding values for a SM-like Higgs of the same mass. For a 125 GeV SM-like Higgs, the dominant decay mode is into a pair of $b$-quarks ($\sim$60\%), followed by $WW$; hence the total width is dominated by the width into $b$ quarks. The largest deviation from mixing effects is expected in the precision measurements of $h \to WW$. This number is reported to be $\mathcal{R}^h_{WW}=1.16^{+0.24}_{-0.21}$ by ATLAS~\cite{Aad:2015gba} and $\mathcal{R}^h_{WW}=0.83\pm 0.21$ by CMS~\cite{Khachatryan:2014jba}. In our analysis we will take a conservative approach of assuming that observational consistency is obtained (that is, the Higgs sector is sufficiently aligned) for $\mathcal{R}^h_{WW}$ between $0.7 - 1.3$. This range will narrow with additional data, and measurements at the level of 10\% are expected at the high luminosity LHC~\cite{ATL-PHYS-PUB-2014-016,CMS:2013xfa}.

\subsection{Vacuum Metastability}

Another important constraint on these parameters comes from vacuum metastability. Large values of the soft stop trilinear coupling $A_t$, required for the Higgs mass and alignment (discussion above), can result in the appearance of charge- and color-breaking minima in the scalar potential of the MSSM. The condition for either these minima to be energetically unfavorable or the tunneling to these minima to have lifetimes longer than the age of the Universe leads to the approximate bound \cite{Blinov:2013fta}
\be \label{metastability1}
A_t^2\lsim \left(3.4 - 0.5\frac{|1-r|}{1+r}\right) m_T^2+60\, m_2^2,
\ee
where $m_T^2=m_{Q_3}^2+m_{u_3}^2, m_2^2=m_{H_u}^2+\mu^2$, and $r=m_{u_3}^2/m_{Q_3}^2$.
In our analysis we assume $m_{Q3}^2 =  m_{u3}^2\equiv M_S^2$, so that $r =  1$. 
Minimization conditions of the Higgs potential give $m_2^2=m_A^2\cos^2\beta+0.5 m_Z^2 \cos(2\beta)$, hence the condition for vacuum metastability can be written as
\be \label{metastability}
A_t^2\lsim 6.8 M_{S}^2+60\, m_A^2\cos^2\beta+30\, m_Z^2 \cos(2\beta).
\ee
It is worth keeping in mind that this is only an approximate bound and depends on several assumptions (see Ref.~\cite{Blinov:2013fta} for details). However, consistency with the above provides a rough guide for the feasibility of the parameter region under investigation. 

\subsection{Direct Detection}
\label{sec:directdetection}

Direct detection possibilities focusing on the $A$-funnel in the MSSM have been studied in Refs.~\cite{Anandakrishnan:2014fia, Hooper:2013qjx, Han:2013gba}. The pseudoscalar $A$ does not mediate spin-independent WIMP-nucleon scattering. Instead this cross section $\sigma_{SI}$ comes  from light and heavy CP-even Higgs boson exchanges in the $t$-channel, facilitated by the Bino-Higgsino mixture of the LSP necessary to obtain the correct relic density. There are also contributions from tree level squark exchange in the $s$-channel and from gluon loops~\cite{Hisano:2010ct, Cheung:2012qy}, but these are negligible when the sfermions are heavy. The cross section then depends only on $M_1,\, m_A,$ $\tan\beta$ and $\mu$. 

For given values of $m_A$ and  $\tan\beta$, requiring the correct relic density and GCE leaves no free parameters, thereby fixing the direct detection cross section. This cross section in our region of interest can be written as approximately \cite{Cheung:2014lqa}
\begin{eqnarray}
\sigma_{SI} \simeq  \frac{ m_Z^2 s_W^2 m_p^2 m_r^2
}{\pi  v^4
  }  N_{11}^4  \left[\frac{\left(\frac{F_u}{t_{\beta }}-F_d
   t_{\beta }\right) }{m_A^2}\left(\frac{N_{14} }{N_{11}}c_{\beta
   }+\frac{N_{13}
   }{N_{11}}s_{\beta
   }\right) -\frac{\left(F_d+F_u\right)}{m_h^2} \left(\frac{N_{13} }{N_{11}}c_{\beta
   }-\frac{N_{14} }{N_{11}}s_{\beta
   }\right)\right]^2,\nonumber\\
   \end{eqnarray}
where $F_u  \sim 0.15,\,F_d  \sim 0.13$~(the up and down type quark content respectively of the nucleon, proton or neutron), $t_\beta=\tan\beta,\, m_N$ is the mass of the nucleon, and $m_r = \frac{m_N m_{\chi}}{m_N+m_{\chi}}$ is the reduced mass. For the correct dark matter relic density obtained via the $A$-funnel, this cross section is generally around $10^{-11}\,$pb \cite{Anandakrishnan:2014fia, Hooper:2013qjx, Han:2013gba, Kim:2002cy}, well below existing bounds from XENON100\,\cite{Aprile:2013doa} and LUX\,\cite{Akerib:2013tjd}, which currently rule out $\sigma_{SI}\gsim 5\times 10^{-10}$ pb. Note that while the annihilation processes that determine the relic density as well as indirect detection signals are $s$-channel and therefore enhanced by the resonance, the direct detection cross-section is mediated by $t$-channel processes and does not receive this enhancement. Such small direct detection cross sections are therefore a generic feature of this region of parameter space. Crucially, this cross section still lies above the neutrino background and is therefore within reach of future detectors, although detection will still be challenging.   

As is well-known, an exception to this generic feature can occur for negative values of the $\mu$ parameter due to destructive interference between the light and heavy Higgs exchange contributions, giving cross sections several orders of magnitude below the neutrino background cross section \cite{Anandakrishnan:2014fia, Han:2013gba}. Such blind spots can in general occur at any dark matter mass, but their appearance in the $A$-funnel framework is more strongly constrained as we also need $m_H\sim m_A\sim 2\,m_{\chi}$. Approximating the up- and down-type quark content in the nucleus as roughly equal, this cancellation condition in the $A$-funnel region can be formulated as approximately~\cite{Anandakrishnan:2014fia} 
\be\label{cancellationcondition}
m_A\sim\left(-2\,\mu \,m_h^2\,\tan \beta\right)^{1/3}\; .
\ee
 With TeV scale values of $\mu$ necessitated by  relic density constraints and $\mathcal{O}(1)$ values of $\tan\beta$ required by collider constraints (see Sec.~\ref{HiggsIntro}), Eq.\,\ref{cancellationcondition} implies that the cancellation can only occur for large $m_A\gsim 650$ GeV, beyond the mass range of interest from the point of view of the GeV excess. Hence all parameter combinations of interest should predict a small but tractable ($\sim10^{-11}\,$pb) direct detection cross section (we will see in the subsequent sections that this in indeed realized, see Fig.\,\ref{fig:direct detection}). 

\subsection{Indirect Detection}

Currently the strongest bounds on the annihilation cross section are given by the Fermi/LAT analysis of 6 years of data on 15 known dwarf galaxies \cite{Ackermann:2015zua}. For $100-300$ GeV dark matter, which is our region of interest, this analysis constrains the annihilation cross-section to be less than $\sim$ a few$\times 10^{-26}$ cm$^3$/s. The cross section required to explain the GCE is also in this region over this mass range (see \cite{Agrawal:2014oha}), hence the dwarf constraints are in some tension with a DM interpretation of the GCE. However, the large uncertainties in the dark matter distribution ($J$-factor) in these dwarf galaxies leave room for compatibility (see Fig.\,8 in Ref.\,\cite{Ackermann:2015zua}). For instance, the 95\%\,C.L. annihilation cross-section exclusion limit for a 100 GeV WIMP annihilating to $b\bar{b}$ is $2.2\times 10^{-26} $cm$^3$/s and has a $1\sigma$ error interval of $[9.0 \times 10^{-27} , 5.6\times 10^{-26}]$ cm$^3$/s, which is compatible with the cross section interval [$3.1\times 10^{-27}, 8.8 \times 10^{-26}$] cm$^3$ needed to fit to the GCE at this mass. A signal was reportedly seen in the new dwarf galaxy candidate Reticulum II \cite{Geringer-Sameth:2015lua}, found in the first year DES data \cite{Bechtol:2015cbp}, consistent with a dark matter of mass $\sim 40-200$ GeV annihilating into $b \bar{b}$ with a cross section $\sigmav \sim 10^{-26}$ cm$^3$/s, although this was later found to be inconsistent with the new {\tt PASS 8} diffuse emission model used to analyze Reticulum II \cite{Drlica-Wagner:2015xua}. Bounds similar to those from the Fermi dwarf observations are also found by the Planck satellite from CMB measurements \cite{Ade:2015xua}. 

Likewise, since DM of interest in this paper annihilates primarily through hadronic channels ($b\bar{b}$ and $t\bar{t}$), this is expected to generate a significant flux of antiprotons. There already exists some tension between models that explain the GCE and derived constraints from antiproton bounds on dark matter annihilation~\cite{Kong:2014haa,Bringmann:2014lpa,Cirelli:2014lwa}. However, calculation of the antiproton flux suffers from significant uncertainties related to the propagation model in the galaxy (see \cite{Cirelli:2014lwa,Cirelli:2013hv,Perelstein:2010gt,Fornengo:2013xda} and references therein), and the GCE can be made compatible with the measured antiproton flux for conservative choices of propagation model parameters.   

Bounds on the dark matter annihilating cross-section into quarks are also obtained by neutrino experiments like IceCube. The most current results from the IceCube-79 experiment exclude $\sigmav \geq 2 \times 10^{-22}$ $\text{cm}^3/\text{s}$ into $b\bar{b}$ at $90\%$ confidence level \cite{Aartsen:2015xej}. This lower limit is $\sim 10^4$ larger than the cross-section required for the GCE \cite{Agrawal:2014oha} and thus irrelevant.

Therefore, no indirect detection results robustly rule out a DM interpretation of the GCE at present, although future measurements, particularly from Fermi-LAT observation of dwarfs, AMS-02 antiproton results, and the CMB could have interesting implications. 
 
 \section{Numerical Results}

Building on the parameter space and constraints described in the previous sections, we present the fits to the GCE excess in this section. We used the following tools for our numerical analysis: the neutralino relic abundance and annihilation cross-section was calculated with {\tt Micromegas-4.1.7}~\cite{Belanger:2014vza}, the MSSM particle spectra were computed using {\tt SuSpect-2.41}~\cite{Djouadi:2002ze}, and the Higgs phenomenology was obtained with {\tt FeynHiggs-2.11.0}~\cite{Hahn:2013ria,Frank:2006yh,Degrassi:2002fi,Heinemeyer:1998np,Heinemeyer:1998yj}. 

For the gamma ray spectrum corresponding to the signal, we follow the approach employed in Ref.~\cite{Agrawal:2014oha} and consider two of the four spectra presented in Fig.\,13 of Ref.~\cite{TheFermi-LAT:2015kwa}\footnote{The first version of our paper used the spectra presented in Ref.~\cite{simonatalk}, and Ref.~\cite{TheFermi-LAT:2015kwa} is the corresponding publication that recently appeared; we have chosen the spectra from Ref.~\cite{TheFermi-LAT:2015kwa} that correspond most closely to the spectra we used in the first version.}, which were derived by fitting the excess over various choices of background as exponentially cut off power laws (see Ref.~\cite{simonatalk,Agrawal:2014oha} for further details).  The four spectra are referred to as spectra (a)-(d) in Ref.~\cite{Agrawal:2014oha}, and just as they do, we pick spectra (b) and (d) for our analysis; spectrum (a) is very similar to what has been studied for light ($m_\chi\lsim 40$ GeV) DM in previous papers and not amenable to the MSSM, whereas spectrum (c) is very similar to spectrum (d) and does not yield any new insight. 

Spectrum (b) corresponds to a fit with OB stars as cosmic ray (CR) sources and a tuned index for pion production within the solar circle (see \cite{simonatalk,TheFermi-LAT:2015kwa}); the analysis in Ref.~\cite{Agrawal:2014oha} found it to be well fit by $75-95$ GeV DM annihilating into $b\bar{b}$ or $\lsim 200$ GeV DM annihilating into $t\bar{t}$. Annihilation into gauge or Higgs bosons were also found to give good fits, but these are irrelevant for our analysis since they are always subdominant channels in the MSSM pseudoscalar resonance scenario. Note that spectrum (b) is also in agreement with other studies performed in Refs.~\cite{Caron:2015wda} and \cite{Bertone:2015tza}, which also found that $175-200$ GeV DM annihilating into $t\bar{t}$ could be compatible with the GCE. Likewise, spectrum (d) corresponds to a fit with OB stars as cosmic ray (CR) sources but with only the intensity of pion production tuned (using pulsars instead of OB stars gives a very similar spectrum); Ref.~\cite{Agrawal:2014oha} found it to correspond to higher mass DM, with $130-165$ GeV DM annihilating into $b\bar{b}$ or $250-310$ GeV DM annihilating into $t\bar{t}$ giving good fits.

In this section, we will perform fits to the two spectra (b) and (d) with the idea of gaining intuition about the range of possibilities that the GCE allows for the MSSM pseudoscalar resonance. We note that the continuous region spanning spectra (b) and (d) could also plausibly explain the GCE for some reasonable background, but do not pursue this direction any further. 

\subsection{Fit Procedure}

The astrophysical information regarding the distribution of dark matter is encoded in the $J$-factor
\begin{equation}
J=\frac{1}{\Delta \Omega}\int_{\Delta \Omega}\int_{l.o.s.}\rho (r)^2 ds d\Omega = \mathcal{J}\times\bar{J}_{can.}, 
\end{equation}  
where $\Delta \Omega$ is the region of interest (ROI), l.o.s. stands for line of sight, and $\rho$ is the dark matter density. $\bar{J}_{can.}~=~2.0~\times~10^{23} \text{GeV}^2/\text{cm}^5$ is the canonical value of the $J$-factor obtained from evaluating the integral with an NFW profile. Following the analyses in Ref.\,\cite{Agrawal:2014oha}, we parametrize the uncertainty in the dark matter density profile with the factor $\mathcal{J}$, which is allowed to vary between $[0.14,4]$. 

The gamma-ray spectrum is computed for the following MSSM parameters:
\begin{itemize}
	
	\item The pseudoscalar mass is allowed to vary over $200$ GeV $\leq m_A \leq 700$ GeV. Below $200$ GeV, we find that the Higgs sector cannot be sufficiently aligned while remaining consistent with bounds from $H/A\rightarrow \tau^+\tau^-$ from the 8 TeV LHC run. We terminate the scan at $700$ GeV since good fits to the GCE (either spectrum (b) or (d)) are not expected for $m_{\chi} \geq 310$ GeV.
	
	\item $\tan\beta$ is scanned over the range $4\leq\tan\beta\leq 10$. Below $\tan\beta=4,$ extremely heavy (multi-TeV) stop masses are required to reproduce the Higgs mass, and large log resummations become important. 
Above $\tan\beta\sim10,\, m_A\lsim 350$ GeV is inconsistent with the LHC $H/A\rightarrow \tau^+\tau^-$ bound. Masses heavier than this do not give good fits to the GCE since $m_A\gsim 310$ GeV ($m_\chi\gsim 165$ GeV) requires annihilation primarily into $t\bar{t}$, but for large values of $\tan\beta$ the leading annihilation channel for the pseudoscalar is into $b\bar{b}$.
	
	\item For given values of $m_A$ and $\tan\beta$, we next scan over $\delta$ (equivalently, $m_\chi$ as shown in Eq.~\ref{eq:delta}) and $\mu$ for points such that  
	\begin{itemize}
	\item the relic density constraint is satisfied: the neutralino makes up all of dark matter ($0.08 \leq \Omega h^2 \leq 0.16$); and 
	\item the annihilation cross section $\sigmav$ is within the $2\sigma$ best-fit annihilation cross-section contours from Ref.~\cite{Agrawal:2014oha}.
	\end{itemize}
	 We scan over $\delta \in [0,0.1]$ in order to stay close to resonance, and over $\mu \in [0.7, 10]$ TeV in order to obtain a mostly bino DM. 
	
	\item Next, we scan over the stop masses $(M_S=m_{Q_3}=m_{u_3})\in[0.7,12.7]$ TeV and the stop trilinear coupling $A_t \in$ $[5,25]$ TeV for points satisfying
	\begin{itemize} 
	\item{$122 \leq m_h \leq 128$ GeV; and }
	\item{alignment in the Higgs sector.}
	\end{itemize}
	 We take the branching ratio to $WW$ normalized to the SM value $\mathcal{R}^h_{WW}$ to be a measure of alignment and select (for each $m_A,\tan\,\beta, \mu, m_{\chi}$ combination) the combination of $M_S$ and $A_t$ that gives $\mathcal{R}_{WW}^h$ closest to $1$ while maintaining $122 \leq m_h \leq 128$ GeV.
	\item  All other MSSM input parameters (gaugino/wino masses, trilinear couplings, slepton/squark masses) are set to $5$ TeV so that they decouple from this analysis.    
\end{itemize}

The goodness of fit is obtained by performing a $\chi^2$ analysis between the gamma-ray spectrum obtained from {\tt Micromegas} and the GCE (Fermi spectra (b) and (d)). For a given MSSM point, the $\chi^2$ is calculated as: 
\begin{equation}
\chi^2 = \sum_k \frac{\left(E_k^2 \frac{dN}{dE_k}(m_\chi ,\mathcal{\bar{J}} \sigmav)- E_k^2\left(\frac{dN}{dE_k}\right)_{obs} \right)^2}{\sigma_k^2},
\end{equation}
where the subscript $k$ runs over the 20 energy bins of the Fermi/LAT measurement \cite{simonatalk}, $dN/dE$ is the gamma-ray spectrum obtained from {\tt Micromegas}, the subscript $obs$ denotes the spectrum consistent with the Fermi excess (\textit{i.e.} spectrum (b) or (d)), $\sigma_k$ denotes the statistical uncertainty \cite{Agrawal:2014oha}, and $\mathcal{\bar{J}}$ is the value of $\mathcal{J} \in [0.14,4]$ that minimizes the $\chi^2 $ value. The $\chi^2$ analysis includes statistical errors, but neglects possible systematic errors from modeling backgrounds near the Galactic Center. 

\subsection{Fit Results}

\begin{figure*}[tb]
	\vskip0.7cm
	\includegraphics[keepaspectratio,width=0.49\textwidth]{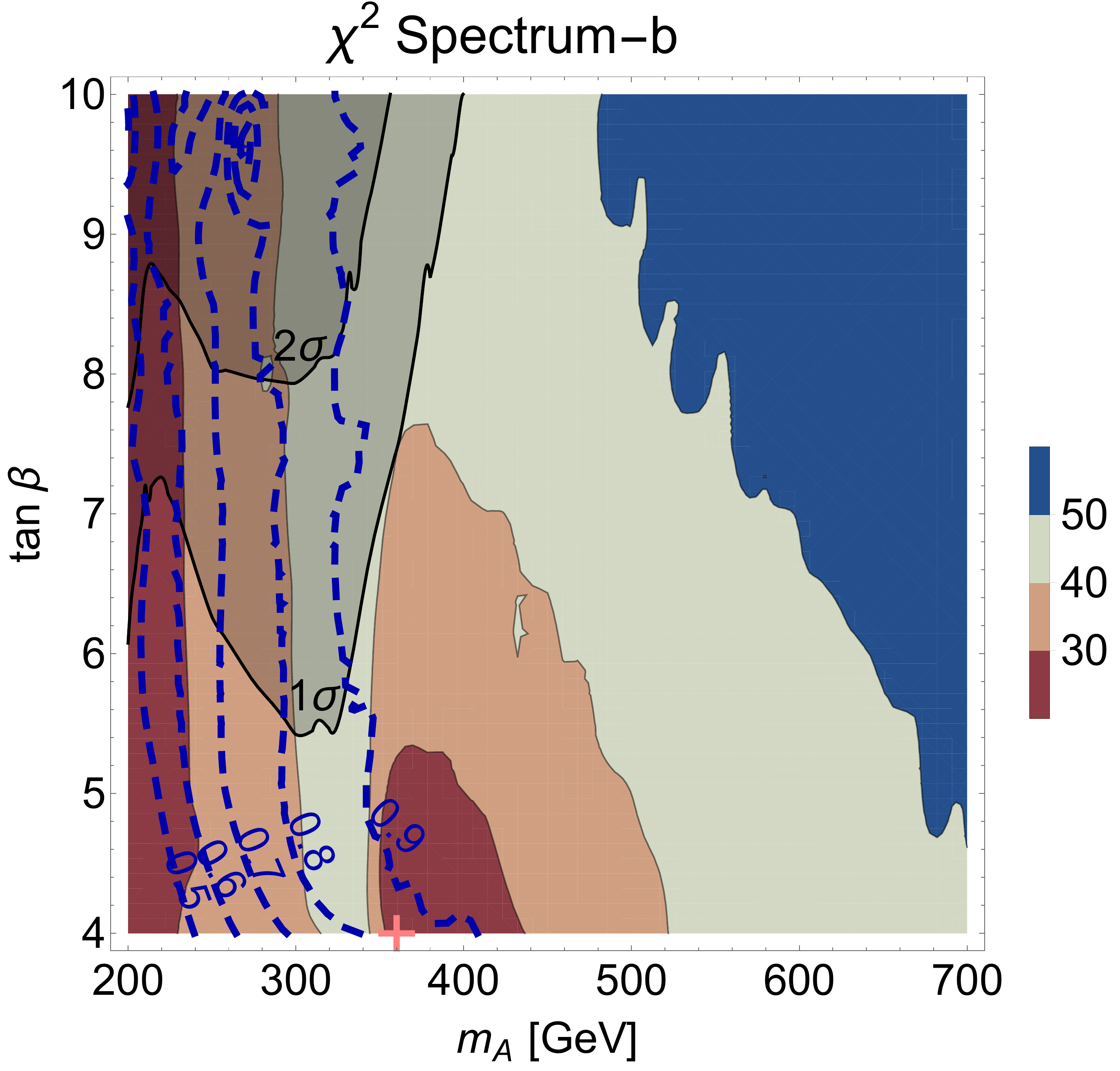}
	\hspace{\stretch{1}}
	\includegraphics[keepaspectratio,width=0.49\textwidth]{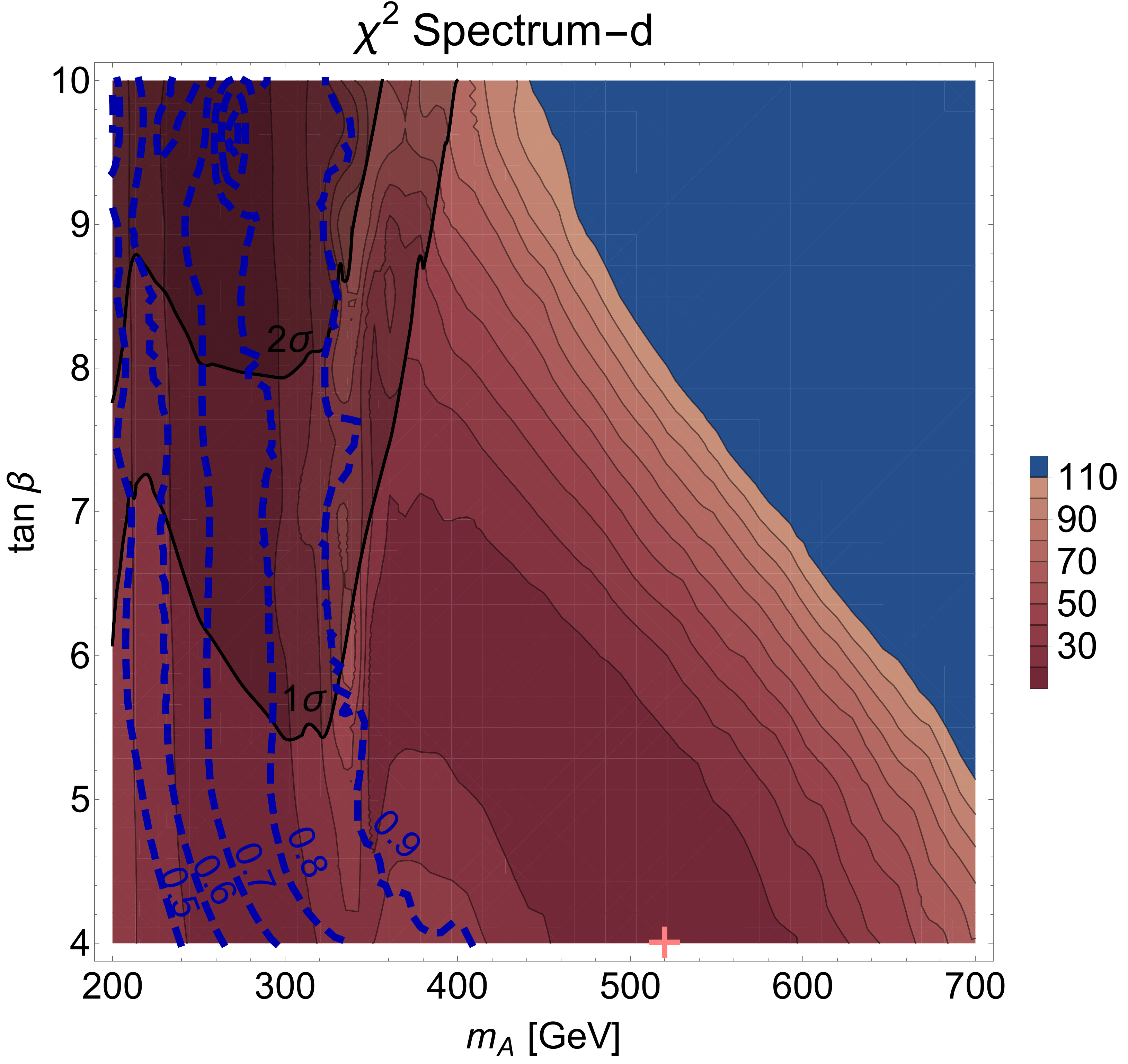}
	\caption{\label{fig:chi2} Contours of $\chi^2$ in the $m_A$-tan\,$\beta$ plane from fitting the gamma-ray spectrum from the MSSM pseudoscalar resonance to Fermi spectrum (b) (left panel) and spectrum (d) (right panel), corresponding to ``OB stars index scaled" and ``OB stars intensity scaled" spectra from Fig.\,13 of Ref.~\cite{TheFermi-LAT:2015kwa} (see Ref.~\cite{simonatalk,Agrawal:2014oha,TheFermi-LAT:2015kwa} for further details). Red (blue) contour regions denote the best (worst) fits. The $\chi^2$ contours are plotted in intervals of 10 in the range $20\leq \chi^2 \leq 110$. The pink crosses denote the lowest $\chi^2$ value in the scan and hence represent the best fit points. Solid black lines mark the 1-$\sigma$ and 2-$\sigma$ exclusion limits (shaded region above the solid black lines excluded) from the negative search results for $H/A \rightarrow \tau^+ \tau^-$ at the 8 TeV LHC run. Dashed blue lines denote contours of the ratio $\mathcal{R}_{WW}^h$; current Higgs data from the 8 TeV LHC favors  $0.7\lsim \mathcal{R}_{WW}^h\lsim 1.3$ (see text for details).
	}
\end{figure*}

\begin{figure}[tb]
	\vskip0.1cm
	\includegraphics[keepaspectratio,width=0.49\linewidth]{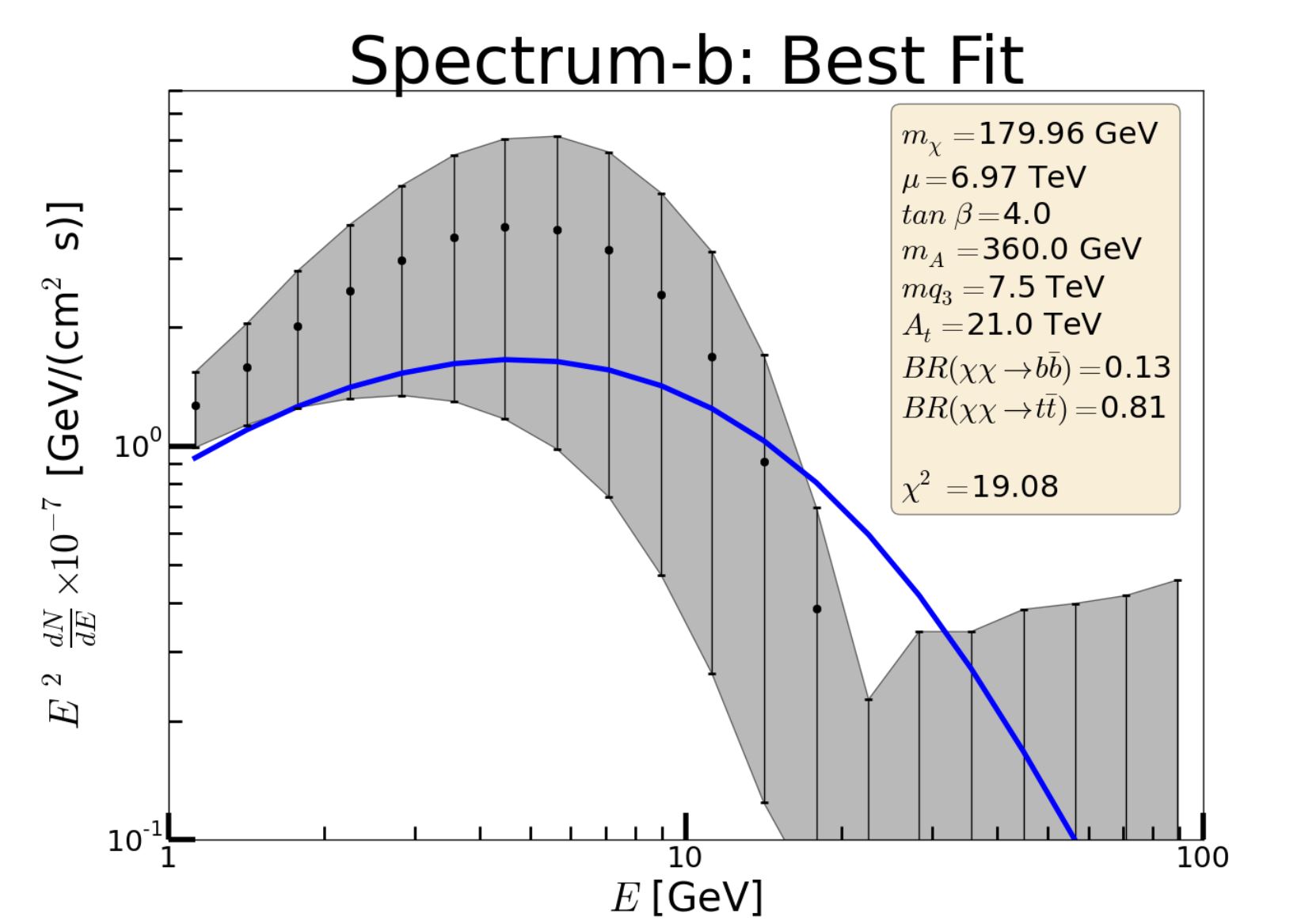}
	\includegraphics[keepaspectratio,width=0.49\linewidth]{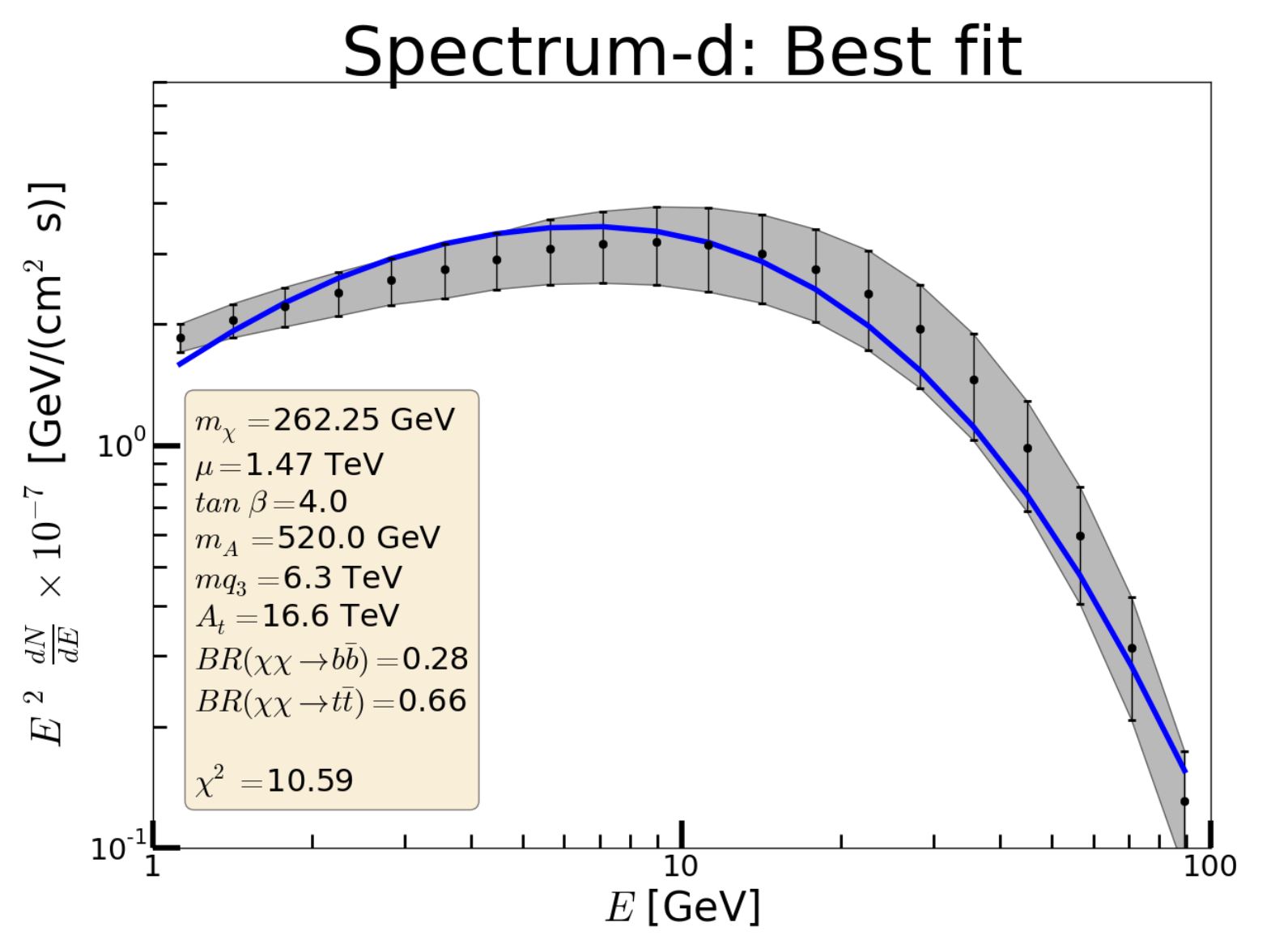}
	\caption{\label{fig:fits} Gamma-ray spectra for the best-fit points corresponding to the pink crosses in Fig.~\ref{fig:chi2}. The gamma-ray spectra from {\tt Micromegas} (blue line) for the best fit points are superimposed on Fermi spectrum (b) and spectrum (d) (black points) on the left and right panels respectively. The gray band denotes statistical uncertainties (from \cite{TheFermi-LAT:2015kwa}). Numerical values of the corresponding MSSM parameters and the leading DM annihilation channels are also listed. The value of the higgs mass, relic density, annihilation cross-section and spin-independent scattering cross-section ($m_h$, $\Omega h^2$, $\sigmav$, $\sigma_{SIp}$) for the best fit point of spectrum-b/d are ($126$ GeV,$ 0.082$, $3.849\times 10^{-26}$ cm$^3$/s, $1.689 \times 10^{-12}$ pb)/($127$ GeV,$0.11$, $3.56\times 10^{-26}$ cm$^3$/s, $4.392 \times 10^{-11}$ pb)}    
\end{figure}

The fits resulting from the above procedure are presented in Fig.~\ref{fig:chi2} as contours of $\chi^2$ in the $m_A$-$\tan\beta$ plane for Fermi spectrum (b) and (d). The pink crosses in each panel denote the points with the best fit to the corresponding spectrum; the gamma-ray spectra of these best fit points are presented in Fig.~\ref{fig:fits} along with the MSSM parameters \footnote{It is worth keeping in mind that the absolute value of $\chi^2$ does not have a proper statistical significance without a full analysis of all uncertainties in the signal and theory prediction.} In Fig.~\ref{fig:chi2} we also include, in solid black lines, the $1$-$\sigma$ and $2$-$\sigma$ bounds from $A/H\rightarrow\tau^+ \tau^-$ searches at the 8 TeV LHC \cite{Khachatryan:2014wca}; points that lie above these curves in the shaded region are inconsistent with these bounds. These $\tau\tau$ searches, however, lose sensitivity at low $\tan \beta$, hence light pseudoscalars can mediate DM annihilations capable of explaining the GCE in this region. The dashed blue lines correspond to contours of $\mathcal{R}^h_{WW}$ as defined in Eq.~\ref{RWW}. $\mathcal{R}_{WW}^h=1$ represents a completely SM-like Higgs, and any mixing with the non-SM Higgs causes deviations. Current Higgs data from the LHC allow for $0.7\lsim \mathcal{R}_{WW}^h\lsim 1.3$, as discussed in Section \ref{HiggsIntro}.  This leads to the requirement of large $\mu$ and hence small couplings~[c.f. Eqs. \ref{N_components} and \ref{achichi}] between $A$ and $\chi$. This generically requires close to resonance conditions $2 m_\chi \approx m_A$ for consistency with both the GCE and relic density.

We found that the $\chi^2$ value did not change significantly between distinct values of  ($\mu, \delta, A_t,$ and $M_S$) for the same $m_A,$ $\tan\beta$. This is expected, since the fit quality is driven by the shape of the spectrum, which is controlled mainly by $\tan\beta$ via the branching ratios, and the position of the peak, which is controlled by $m_A (\approx 2 m_\chi)$. Although the fit should also depend on the signal strength, which is controlled by $\mu$ and $\delta$ via the annihilation cross section and relic density, the freedom in choosing $\mathcal{J} \in [0.14,4]$, which essentially rescales the signal strength, smears out this dependence. In our region of interest, we find that $\delta\lsim 0.04$ while $M_S, A_t,$ and $\mu$ all take multi-TeV values; we present contour plots of these parameters in Fig.~\ref{fig:Param} in Appendix \ref{appendix1}. The condition for vacuum metastability, Eq.\,\ref{metastability}, is also found to be satisfied in most parts of the parameter space allowed by the 8 TeV LHC $A/H\rightarrow\tau^+ \tau^-$ bounds (see Fig.~\ref{atmqratio} in Appendix \ref{appendix1}).

From the left panel of Fig.~\ref{fig:chi2}, the best fit regions to Fermi spectrum (b) appear to be separated into two distinct islands. The $m_A\lsim 250$ GeV region has relatively low $\chi^2$ for all values of tan$\,\beta$. In this region, annihilation into top quark pairs is kinematically forbidden, so the dominant annihilation channels is $b\bar{b}$ for all values of $\tan\beta$. Recall that an approximately  100 GeV DM particle annihilating into $b\bar{b}$ can fit the GCE \cite{Agrawal:2014oha}; this region reflects this behavior. However, we see that this region is incompatible with the 8 TeV LHC $A/H\rightarrow\tau^+ \tau^-$ bounds and/or the Higgs data (that is, $\mathcal{R}_{WW}^h \lsim 0.7$ in this region, signaling that the heavier CP-even scalar is so light that alignment does not work well). A second island opens up at $350$ GeV $\lsim m_A\lsim 450$ GeV, when annihilation into $t\bar{t}$ becomes kinematically feasible, and $\tan\beta\lsim 6$. This is consistent with Ref.~\cite{Agrawal:2014oha} finding a $\sim200$ GeV DM annihilating into $t\bar{t}$ providing a good fit to spectrum (b). Note that the best fit point occurs at the lowest allowed value of $\tan\beta$(=4) in our scan, where the coupling of $A$ to top quarks is the largest. The fit deteriorates as $\tan\beta$ gets larger, as the branching ratio into $b\bar{b}$ gets larger due to the $\tan\beta$ enhancement of the $Abb$ coupling. This region is also compatible with Higgs data as $\mathcal{R}_{WW}^h\gsim 0.7$, and safe from the current $A/H\rightarrow\tau^+ \tau^-$ bounds. Beyond this island, the fit deteriorates rapidly as $m_A$ and/or tan$\,\beta$ are increased. 

Similar patterns are observed for the fit to spectrum (d). A small region of good fit exists at $m_A\sim 300$ GeV and low $\tan\beta$, safe from the $A/H\rightarrow\tau^+ \tau^-$ bounds and borderline compatible with Higgs data. Again, DM in this region annihilates dominantly to $b\bar{b}$ since $t\bar{t}$ is kinematically forbidden, and this observation is compatible with Ref.~\cite{Agrawal:2014oha}, where DM with mass $130-165$ GeV annihilating into $b\bar{b}$ was found to give good fits to the spectrum. A second region with better fits is again observed for larger $m_A$ once decay into $t\bar{t}$ opens up. This regions roughly spans $450$ GeV $\lsim m_A\lsim 600$ GeV and tan$\,\beta\lsim 8$, and appears to correspond to the $250-310$ GeV DM annihilating into $t\bar{t}$ region reported in Ref.~\cite{Agrawal:2014oha} as a good fit to spectrum (d). Similarly to spectrum (b), the best fit occurs for small values of tan$\,\beta\ $: tan$\,\beta\sim 4.0$. This suggests that a DM candidate that annihilates significantly into $t\bar{t}$ with $BR(\chi \chi \rightarrow t \bar{t})=0.66$ at the best fit point) provides the best fit to spectrum (d). This can be confirmed by comparing the shape of the spectrum in Fig.\,\ref{fig:fits}, right panel, which fits the shape of Fermi spectrum (d) quite well. Finally, the fit deteriorates for larger $m_A$ and tan$\,\beta$ values and we do not expect any good fits beyond the region shown in the plot.

\subsubsection{Fit to a Modified Spectrum}
\begin{figure*}[tb]
	\vskip0.7cm
	\includegraphics[keepaspectratio,width=0.49\textwidth]{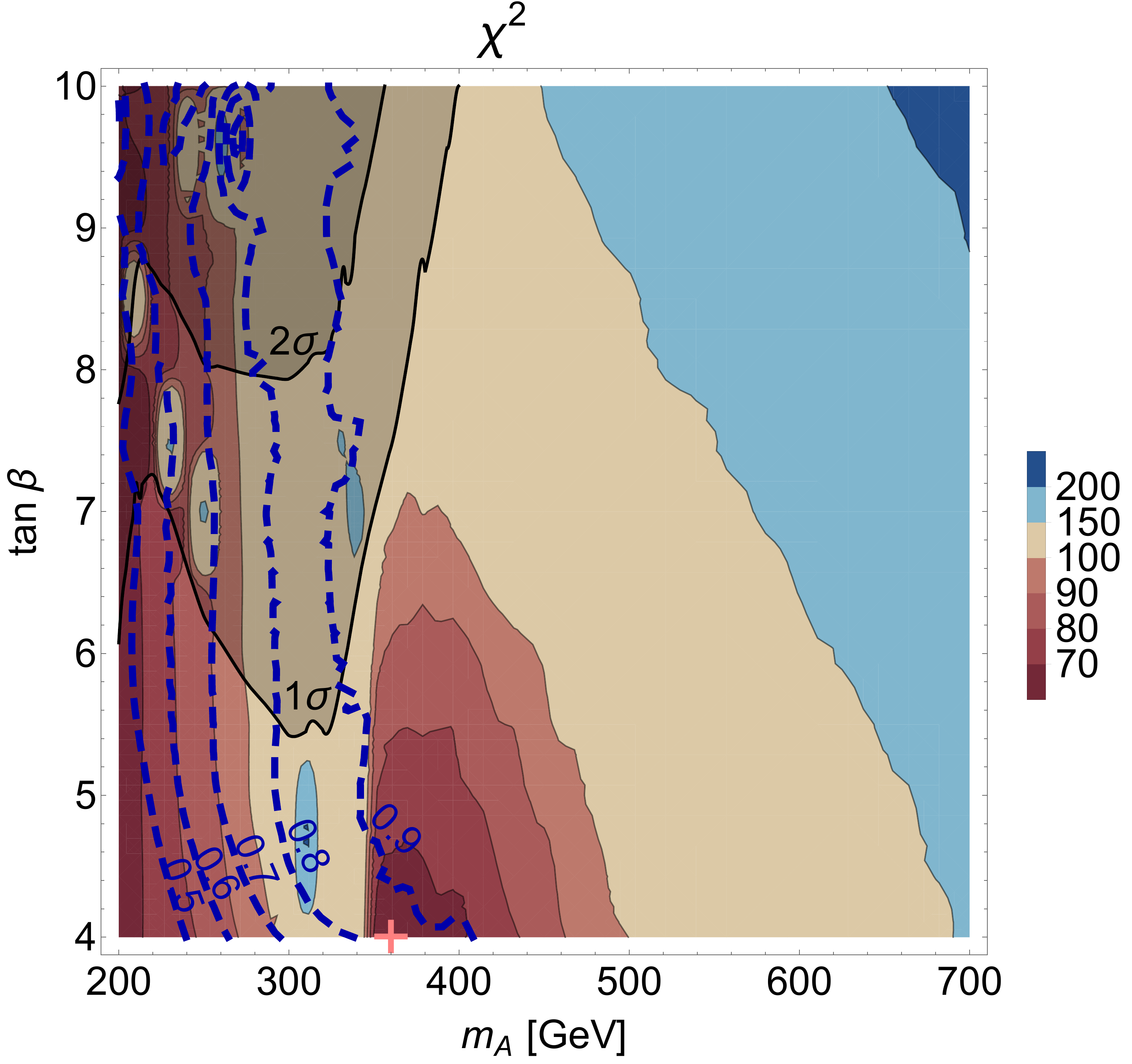}
	\caption{\label{fig:chi2modified} Contours of $\chi^2$ in the $m_A$-tan\,$\beta$ plane from fitting the gamma-ray spectrum from the MSSM pseudoscalar resonance to the ``pulsars index scaled" spectrum from Fig.\,18 of Ref.~\cite{TheFermi-LAT:2015kwa}. Red (blue) contour regions denote the best (worst) fits. Black and blue contours are as in Fig.\,\ref{fig:chi2}. 
	}
\end{figure*}

So far, we performed fits to spectra (b) and (d) as defined in Ref.~\cite{Agrawal:2014oha}, corresponding to the ``OB stars index scaled" and ``OB stars intensity scaled" spectra from Fig.\,13 of Ref.~\cite{TheFermi-LAT:2015kwa}, which were obtained by modeling the excess with an NFW profile with a single power law with an exponential cutoff. This mimics what is expected of a dark matter source, and serves the purpose of demonstrating how the preferred theory parameter space changes for two different choices of interstellar emission models of the background (matching the philosophy in Ref.~\cite{Agrawal:2014oha}). However, Ref.~\cite{TheFermi-LAT:2015kwa} also finds significantly better fits to the excess if more freedom is allowed in the fit -- in particular, if the spectrum of the NFW profile is modeled with a power-law that is allowed to vary per energy band over the 1 - 100 GeV range; the resulting spectra for various choices of interstellar emission models are presented in Fig. 18 of Ref.~\cite{TheFermi-LAT:2015kwa}. In order to study how the MSSM fit is affected if the latter is used, we performed a similar fit (as described above) to the ``pulsars index-scaled" spectrum from Fig. 18 of Ref.~\cite{TheFermi-LAT:2015kwa}; the result is shown in Fig.\,\ref{fig:chi2modified}. We find that the overall fit quality worsens due to the tail of the spectrum, but the best fit regions in the MSSM parameter space still closely match those from the fit for spectrum (b) (see Fig.\,\ref{fig:chi2} (left)); consequently, the theoretical implications from fitting to spectrum (b) (discussed below) will also apply in this case.  

\section{Predictions for the LHC and Direct Detection Experiments}

\subsection{LHC Prospects}

\begin{figure*}
	\includegraphics[keepaspectratio,width=0.45\textwidth]{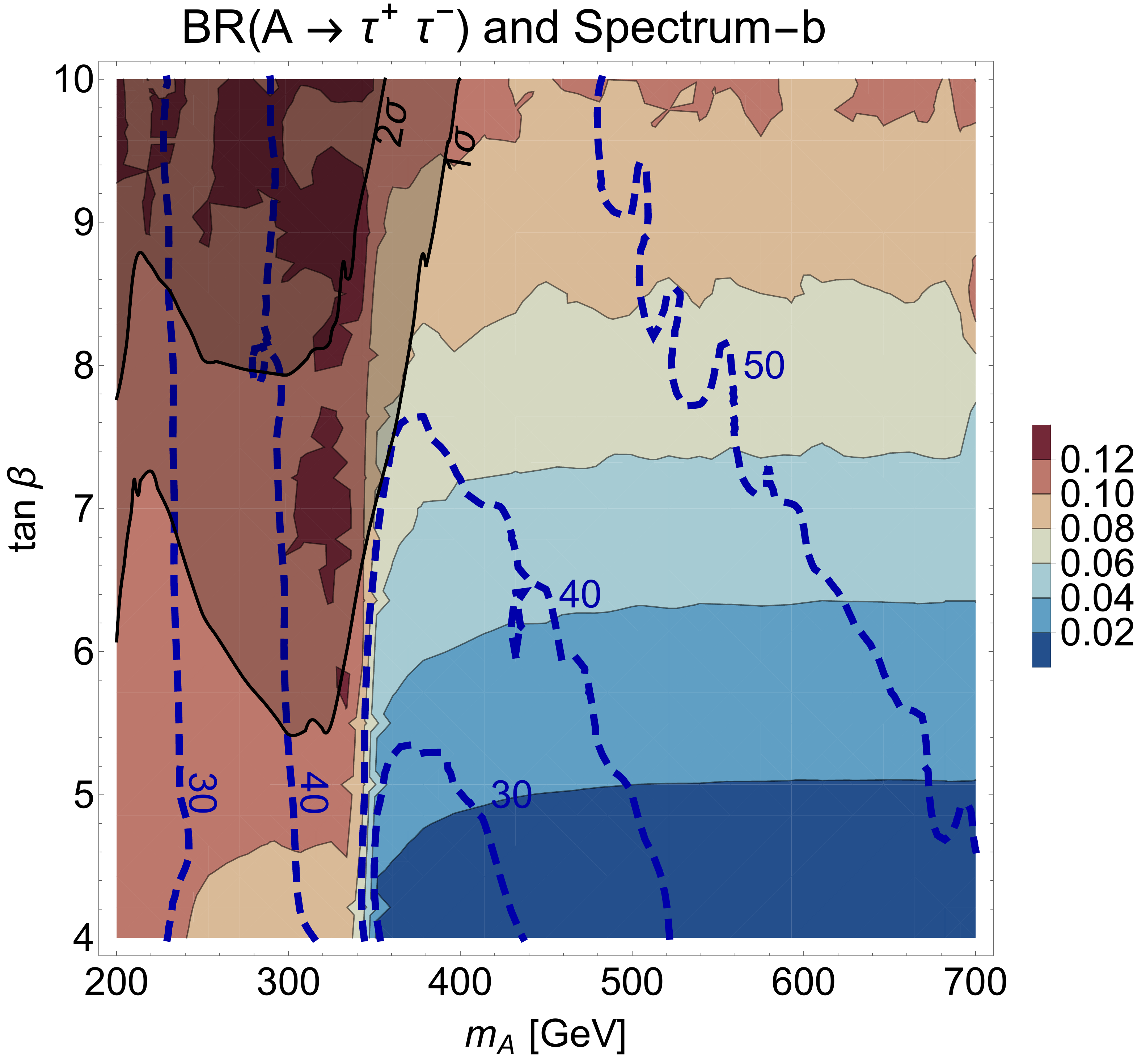}
	\includegraphics[keepaspectratio,width=0.45\textwidth]{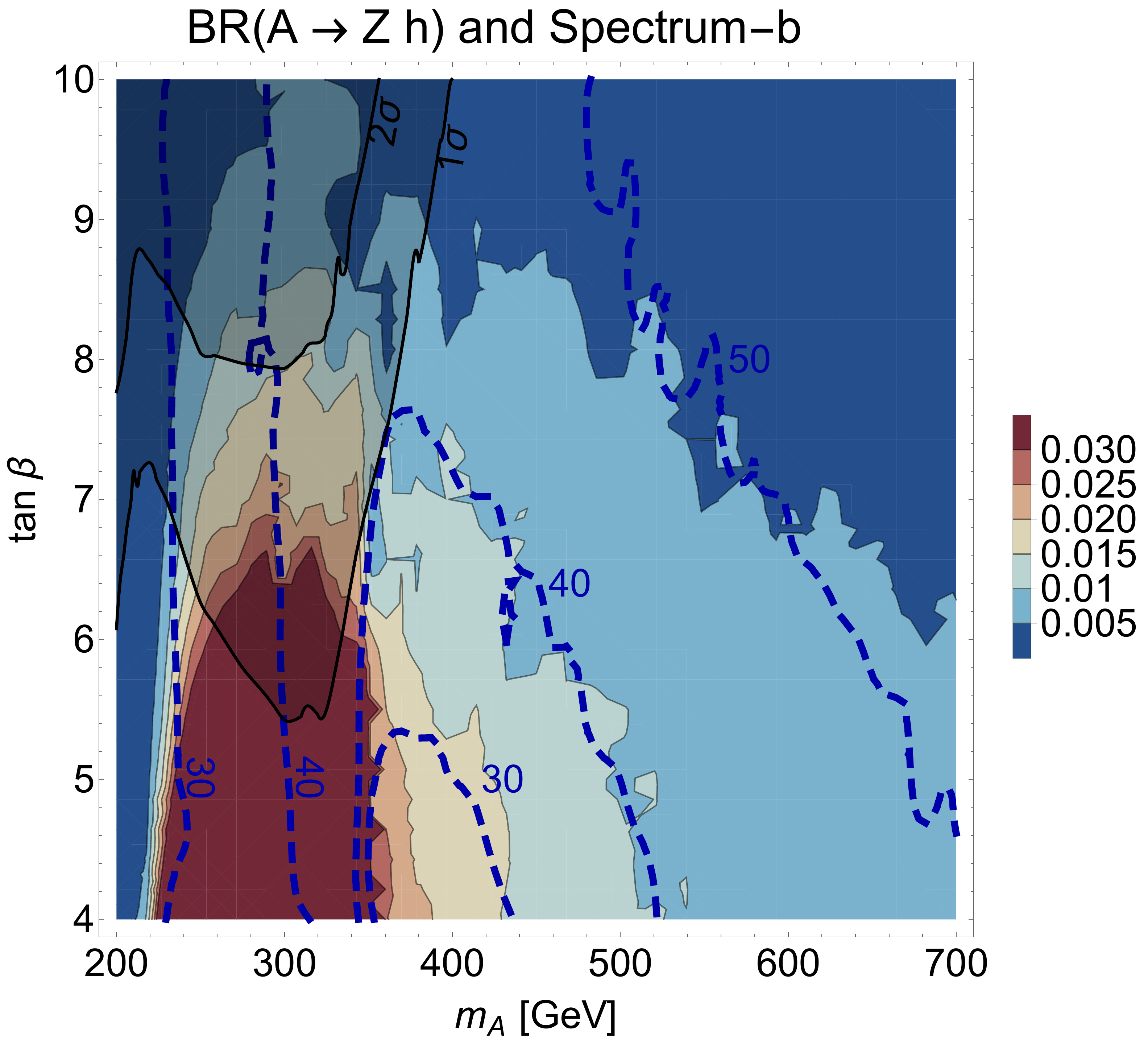}
	\hspace{\stretch{1}}
	\includegraphics[keepaspectratio,width=0.45\textwidth]{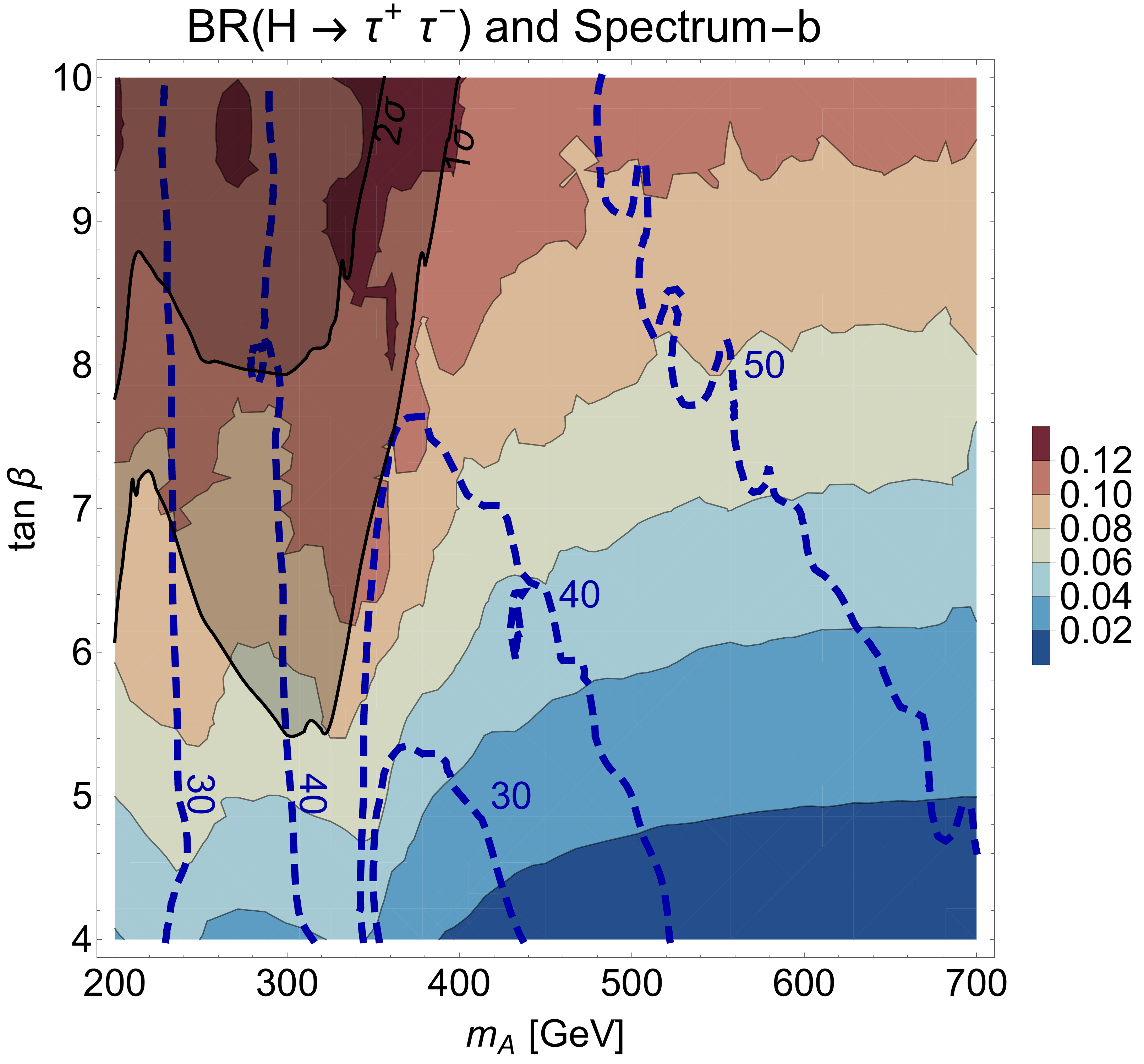}
	\includegraphics[keepaspectratio,width=0.45\textwidth]{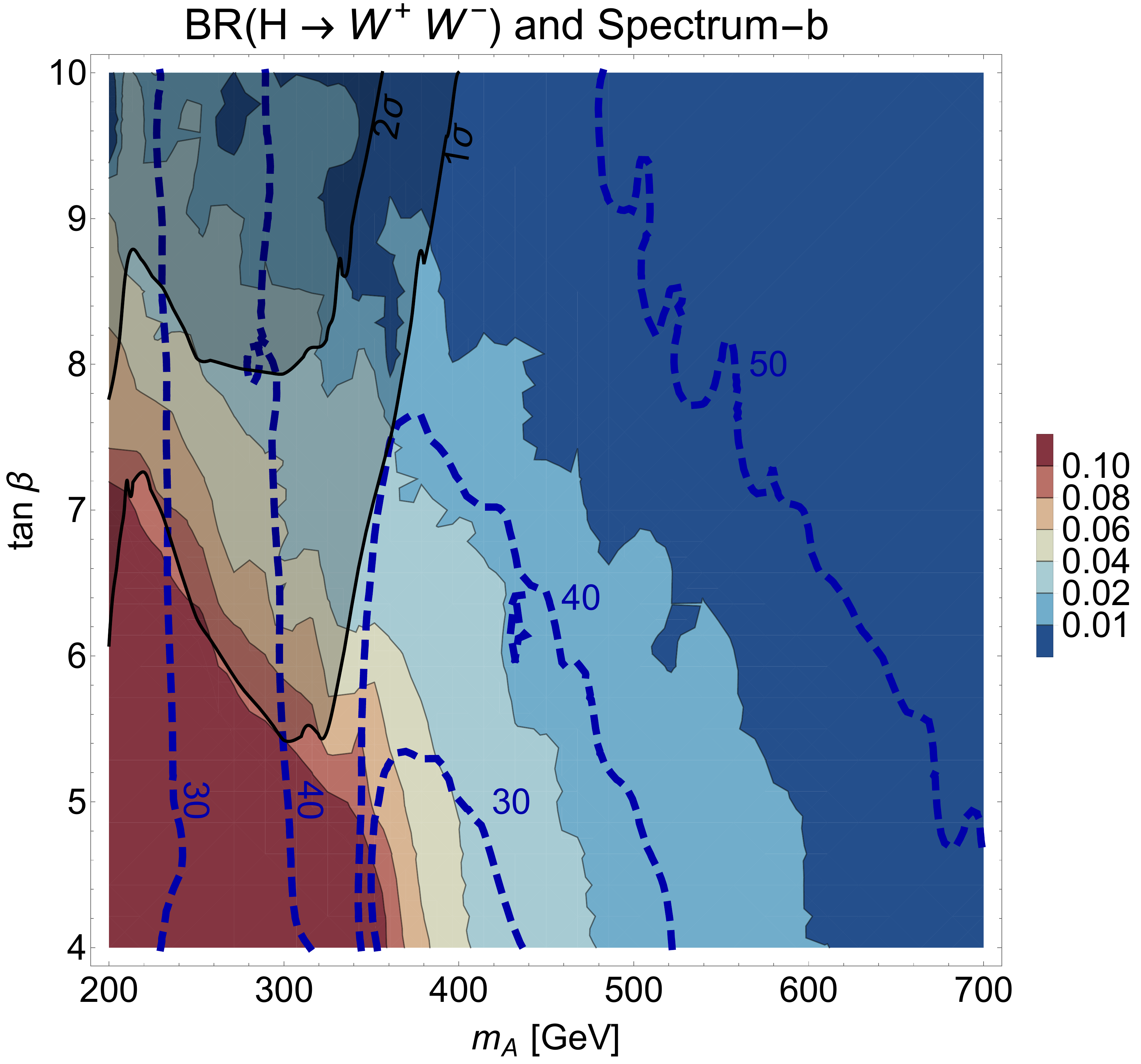}
	\includegraphics[keepaspectratio,width=0.45\textwidth]{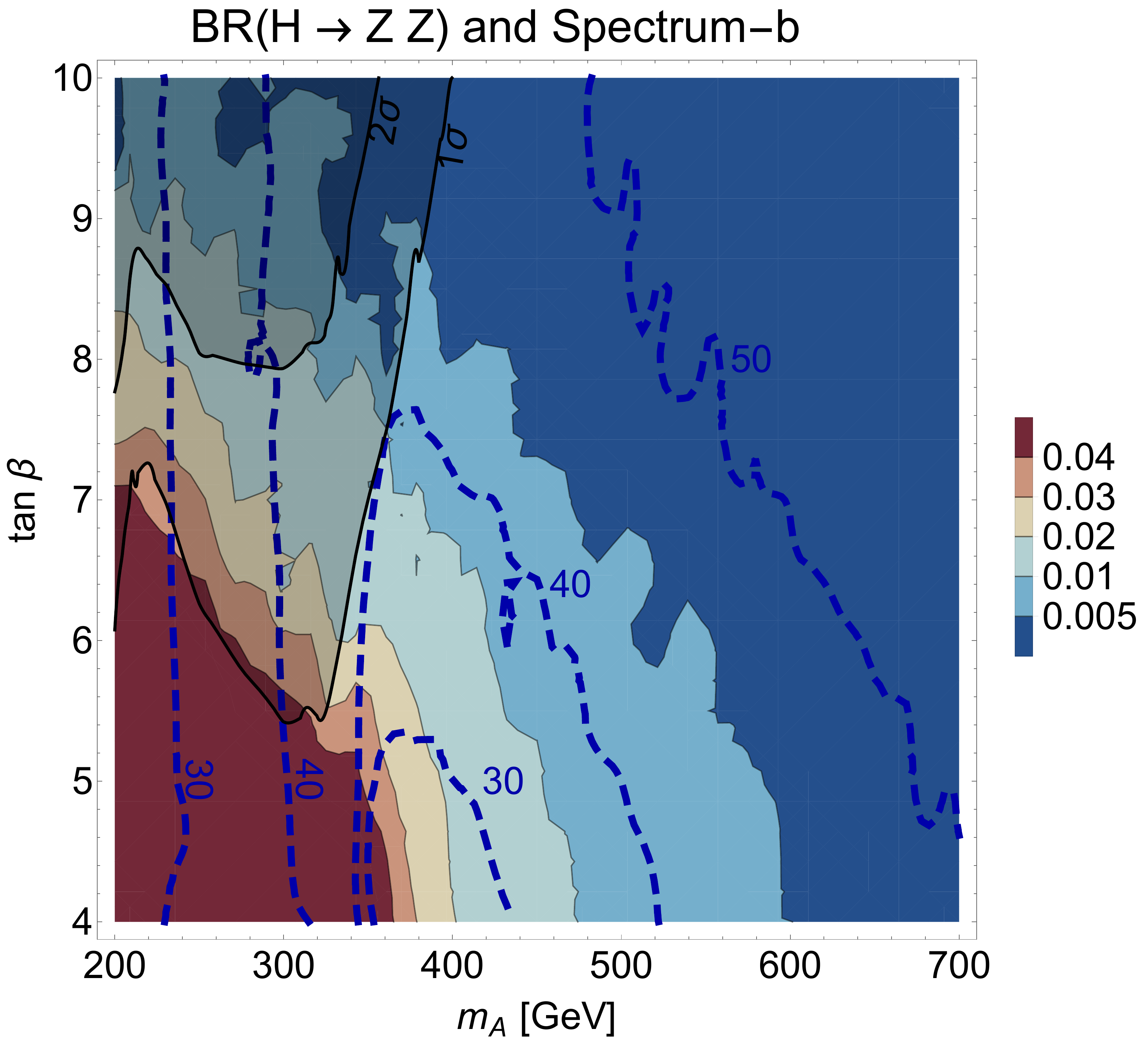}
	\includegraphics[keepaspectratio,width=0.45\textwidth]{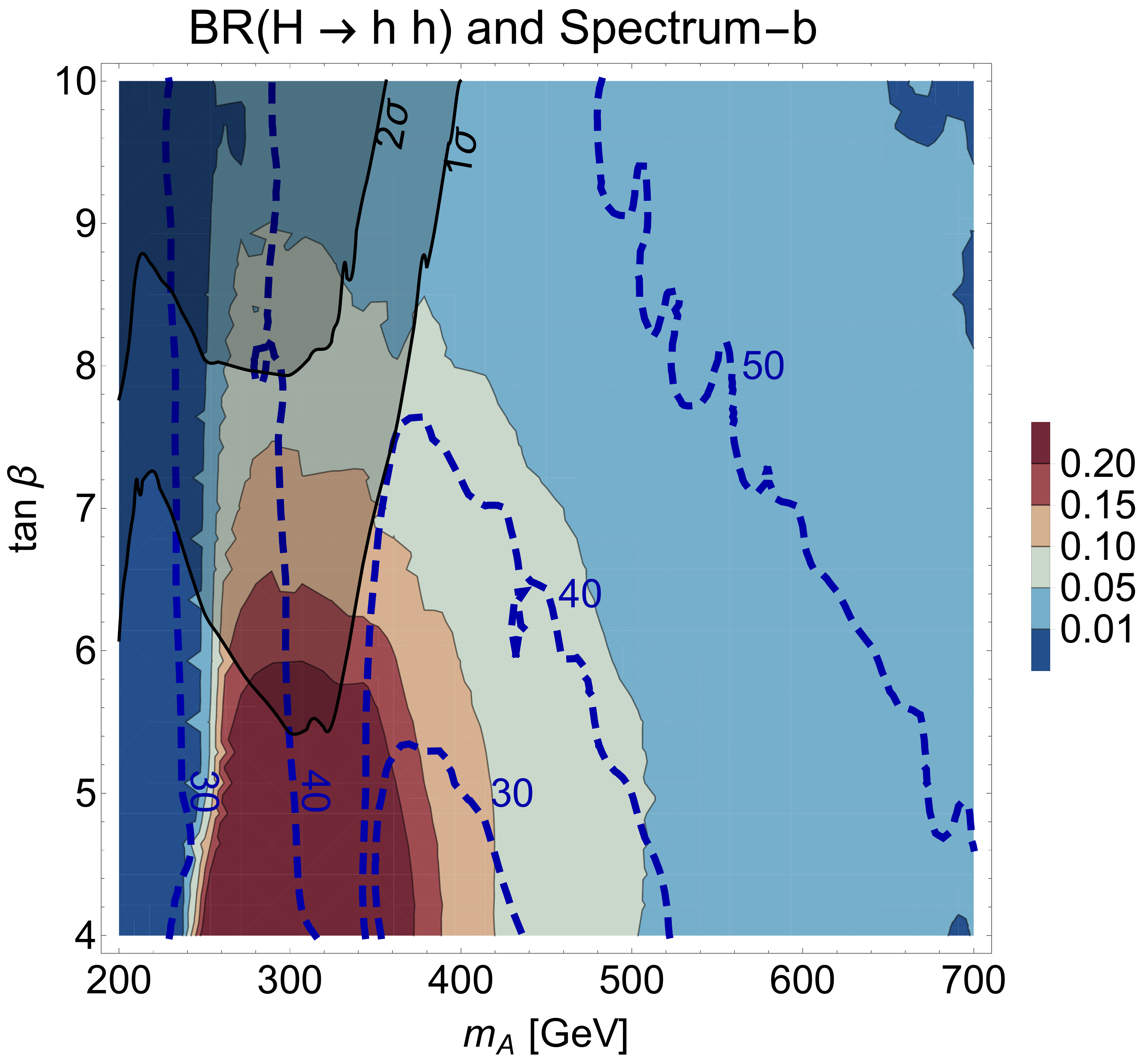}
	\caption{\label{fig:BRb}
		Predictions for LHC. {\it Top}: $BR(A\rightarrow \tau^+\ \tau^-$) and $BR(A\rightarrow Z\ h$). {\it Middle}: $BR(H\rightarrow \tau^+ \tau^-$) and $BR(H\rightarrow W^+ W^-$). {\it Bottom}: $BR(H\rightarrow Z Z$) and $BR(H\rightarrow h h$). Dashed blue lines show $\chi^2$ values from fitting the GCE to spectrum-b, as seen in Fig.~\ref{fig:chi2}. The colored contour regions (and bar on the right) are each plot's respective branching ratio values. Shaded regions labelled 1-$\sigma$ and 2-$\sigma$ are the $A/H \rightarrow \tau^+ \tau^-$ exclusion limits.
		}
\end{figure*}

\begin{figure}
	\includegraphics[keepaspectratio,width=0.45\textwidth]{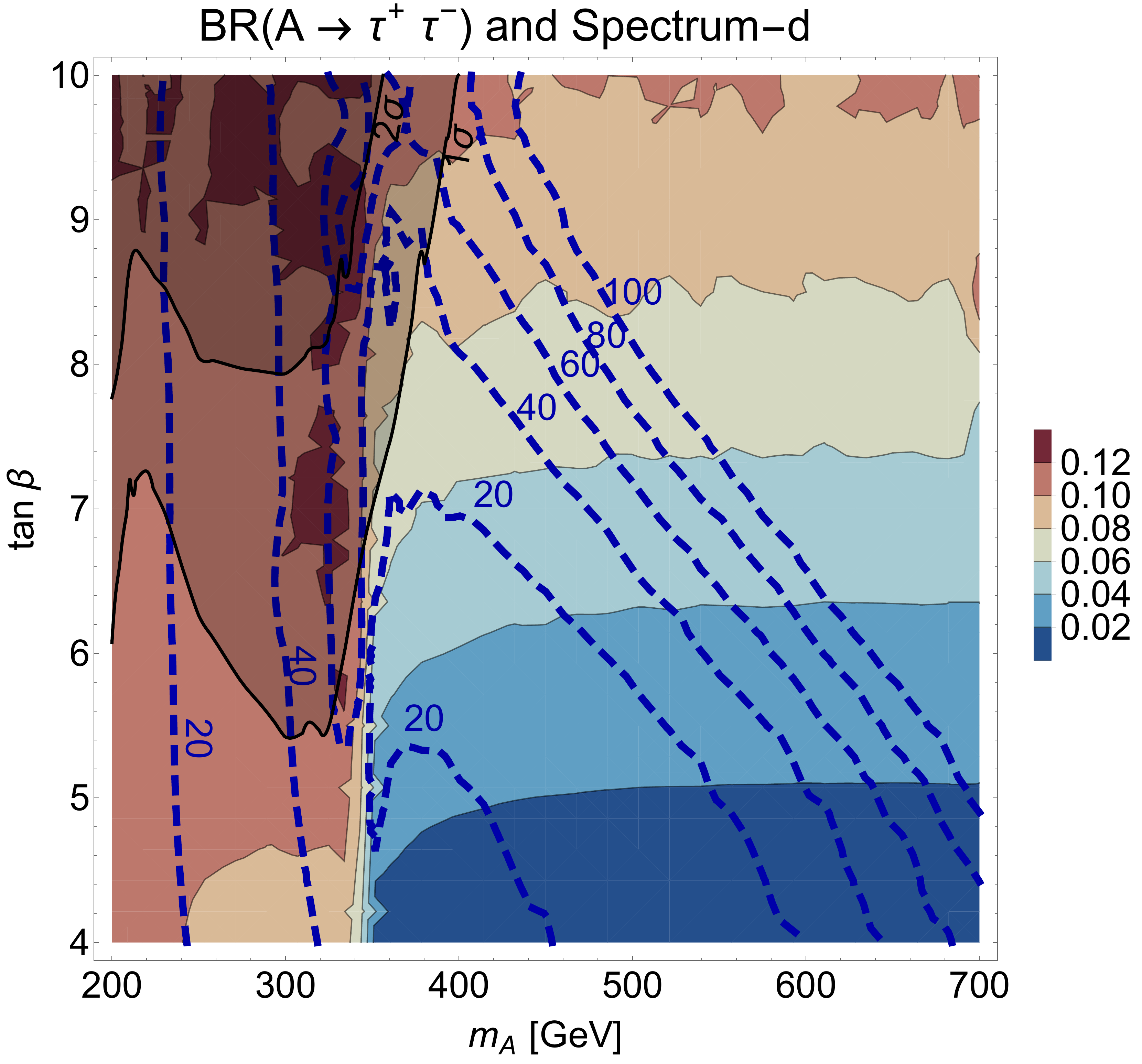}
	\includegraphics[keepaspectratio,width=0.45\textwidth]{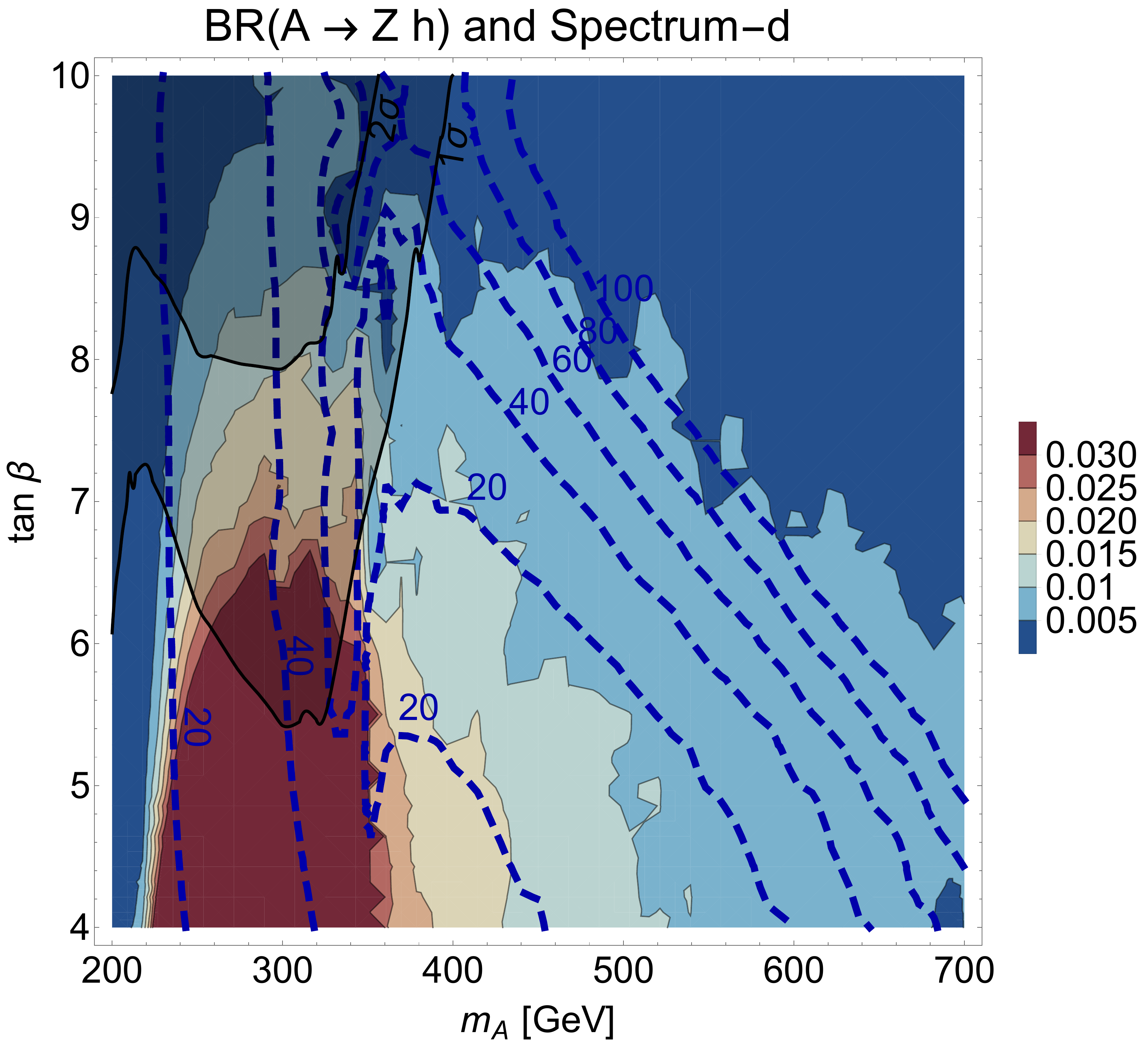}
	\hspace{\stretch{1}}
	\includegraphics[keepaspectratio,width=0.45\textwidth]{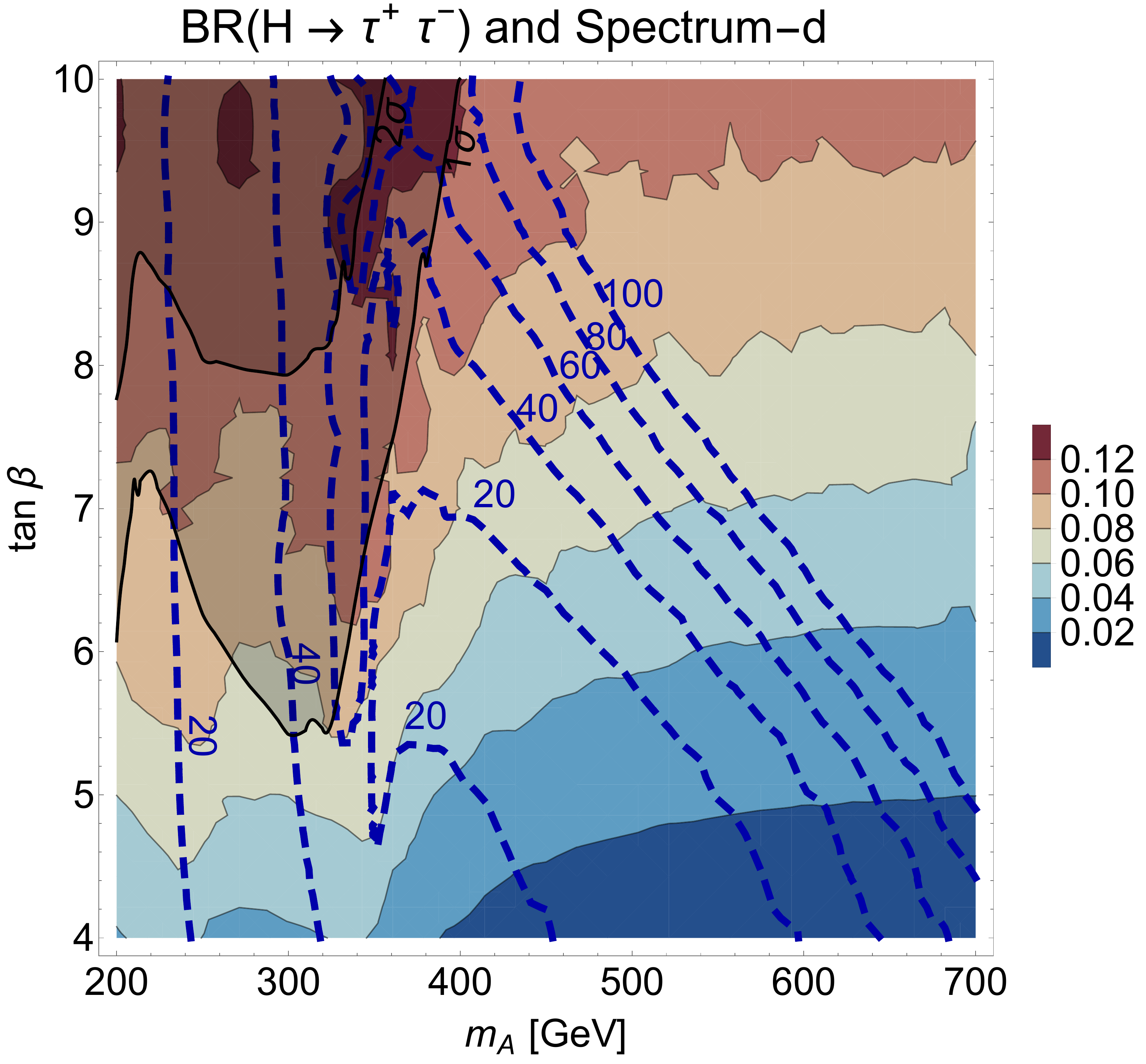}
	\includegraphics[keepaspectratio,width=0.45\textwidth]{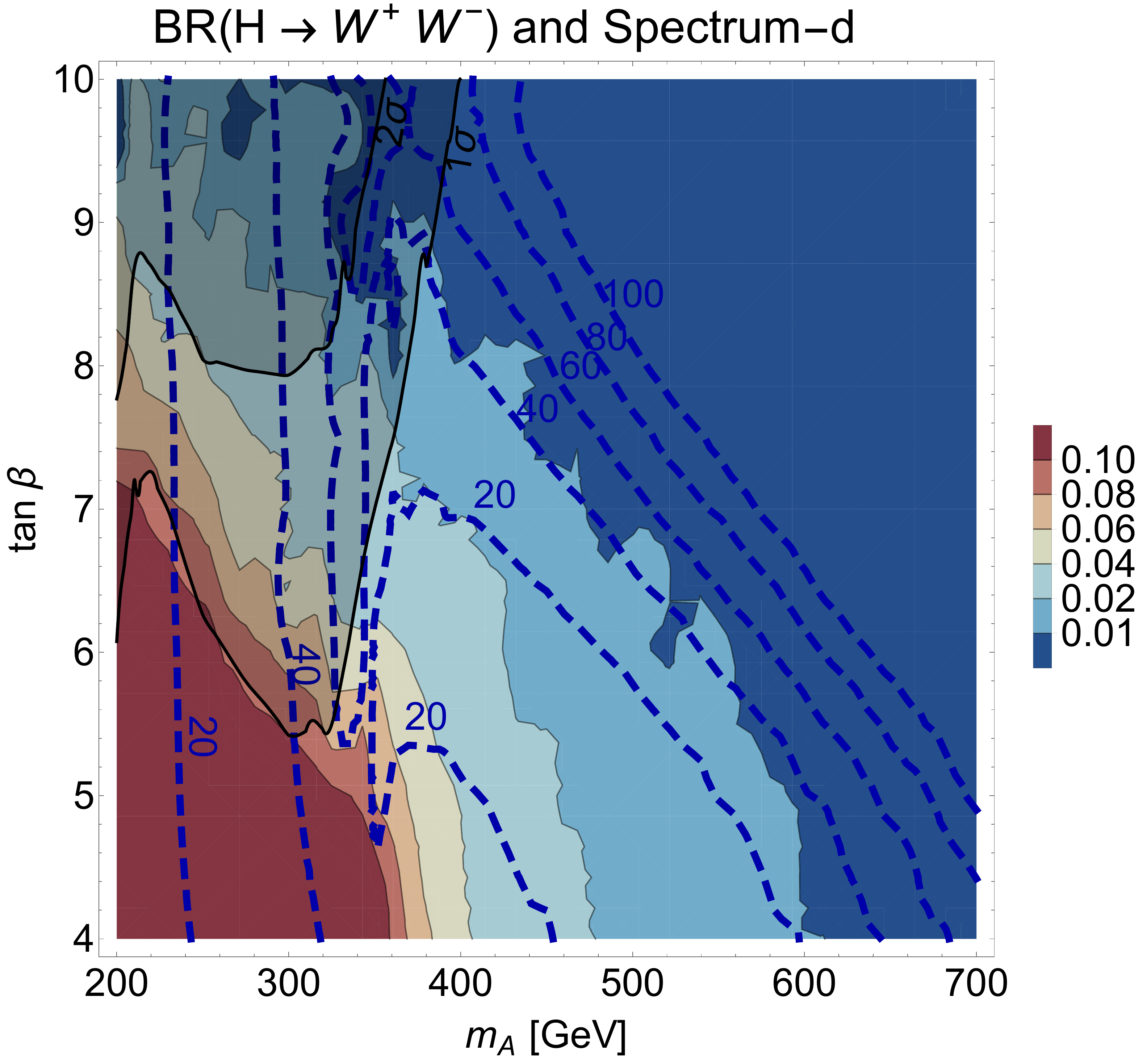}
	\includegraphics[keepaspectratio,width=0.45\textwidth]{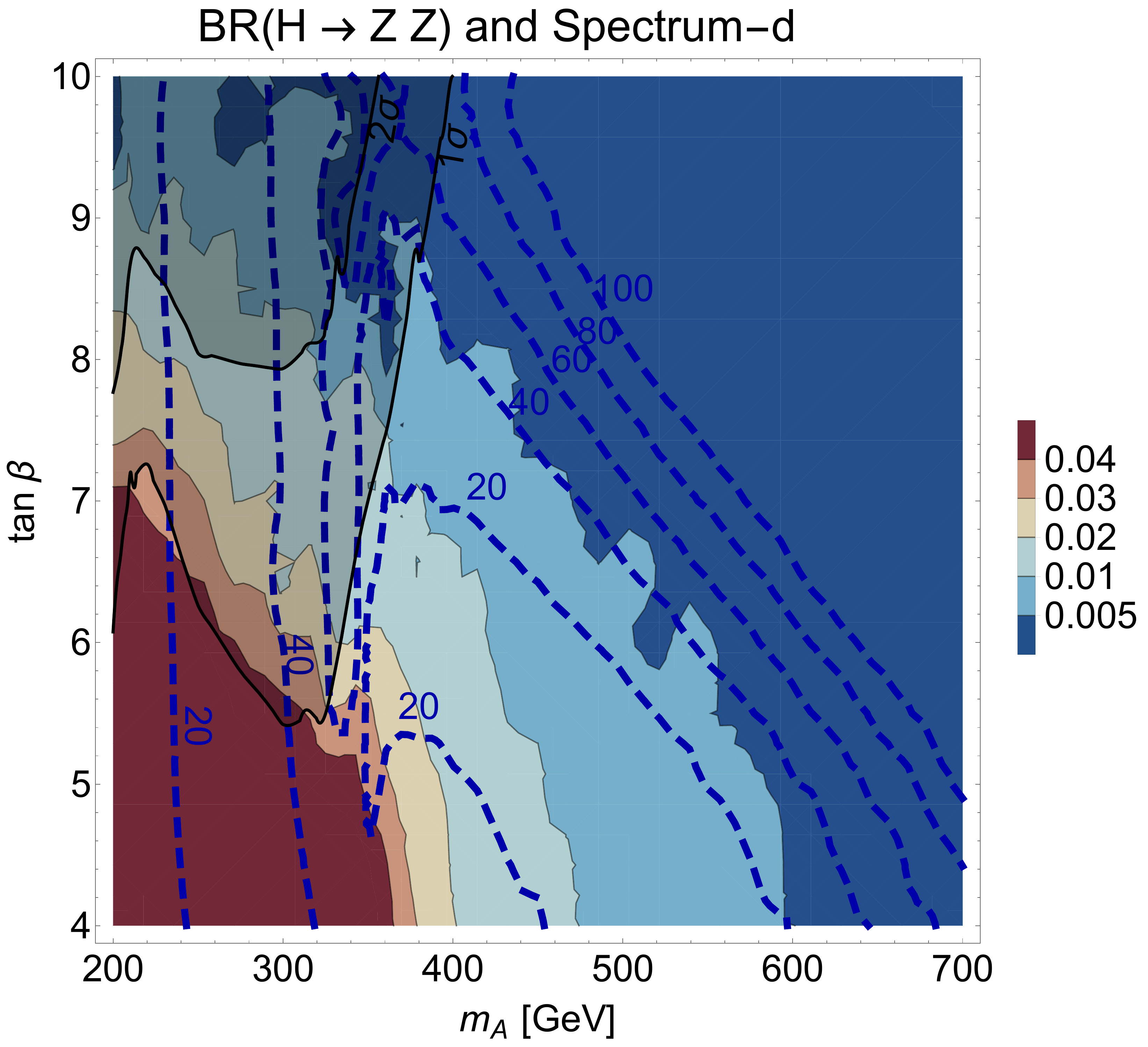}
	\includegraphics[keepaspectratio,width=0.45\textwidth]{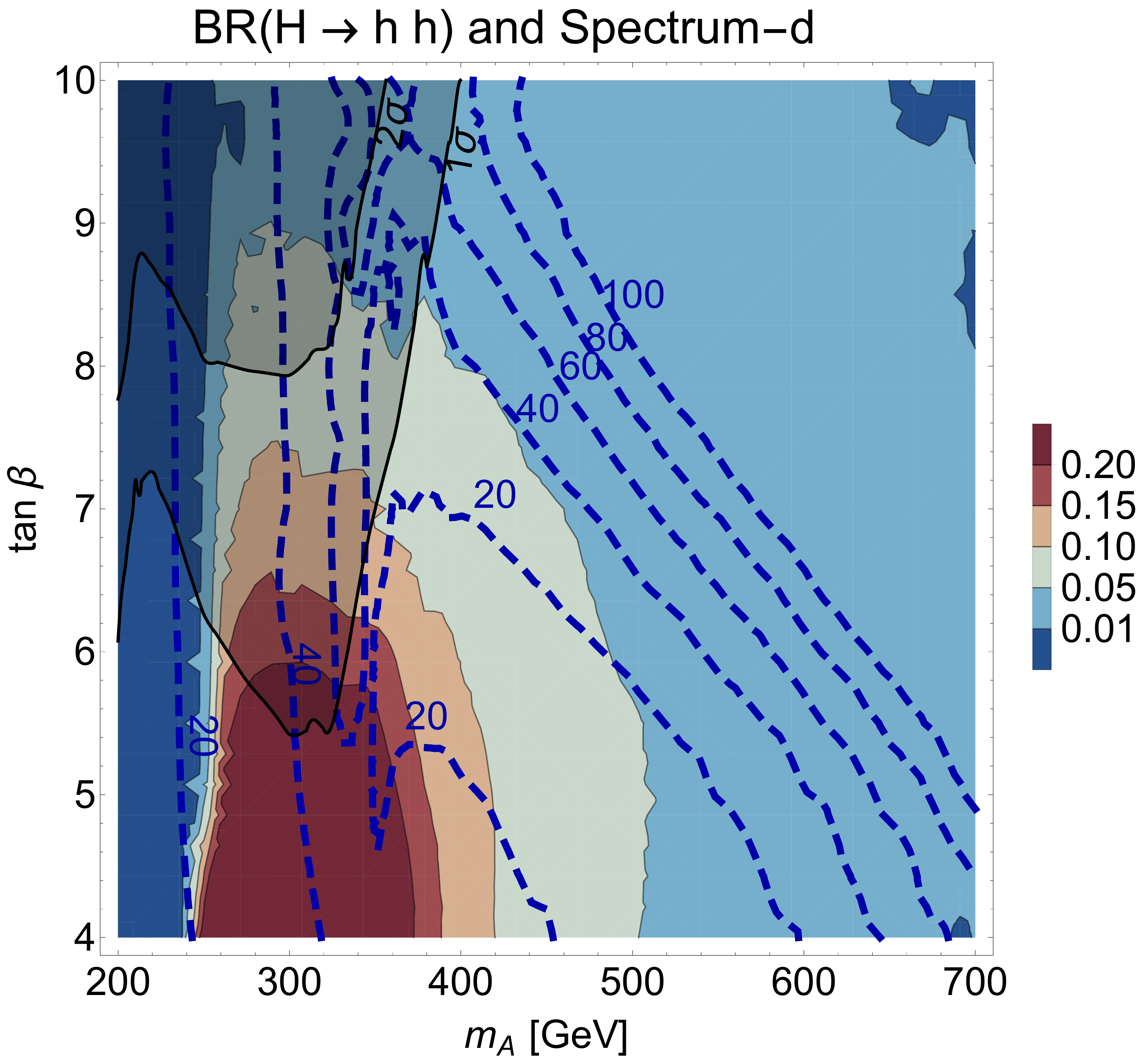}
	\caption{\label{fig:BRd}
		Same as Fig.~\ref{fig:BRb} but with $\chi^2$ values from fitting the Galactic Center excess to spectrum-d, as shown in Fig.~\ref{fig:chi2}.}

\end{figure}

There are several projections for the 14 TeV LHC provided by the CMS and ATLAS collaborations for heavy Higgs searches in \cite{ATLAS-PHYS-PUB-2013-016,CMSPASFTR-13-024}. There are also several theoretical studies showing the hypothesized sensitivity of the 14 TeV LHC in the $m_A-\tan\beta$ plane due to different search channels, for example Ref. \cite{Djouadi:2015jea}. In Figs.~\ref{fig:BRb} and \ref{fig:BRd} we show the interplay between possible interesting signatures for $H/A$ searches at the LHC and the GCE best fit regions in the $m_A-\tan\beta$ plane, plotting contours of various branching ratios of interest for $H/A$ searches at the LHC. To highlight the regions of interest, we overlay the $\chi^2$ values from Fig.\,\ref{fig:chi2} as dashed blue lines for Fermi spectrum (b) in Fig.~\ref{fig:BRb} and spectrum (d) in Fig.~\ref{fig:BRd}. The gray shaded regions denote the current LHC exclusion limits from searches for $H/A\to \tau^+\tau^-$ at the 8 TeV LHC (1-$\sigma$ and 2-$\sigma$ as labeled).   
In both figures, the two panels in the top rows show the branching ratios of the CP-odd Higgs: $A \rightarrow \tau^+ \tau^-$ (left) and $A\rightarrow Z h$ (right). The lower four panels display the branching ratios for $H\rightarrow \tau^+ \tau^-$ (middle left),  $H\rightarrow W^+ W^-$ (middle right),  $H\rightarrow Z Z$ (lower left) and $H\rightarrow h h$ (lower right). 

The top row shows that both $BR(A\to \tau\tau)$ and $BR(A\to Zh)$ are a few percent throughout the parameter region of interest, with the former always comparable to or larger (in some cases, by more than an order of magnitude). We can understand this behavior by noting that due to the close to alignment conditions, the $AZh$ coupling is very suppressed. Hence, despite the $\tan\beta$ enhancement of the gluon fusion production of $A$, we find that the rates for $A \to Zh$ are at least 2 orders of magnitudes smaller than the current exclusion limits~\cite{Aad:2015wra,Khachatryan:2015lba} and therefore unlikely to be probed even at the high luminosity LHC \cite{ATLAS-PHYS-PUB-2013-016,CMSPASFTR-13-024}.  Due to the absence of any other relevant decay modes, the decays to down-type fermions will still be the dominant decay modes and offer the best prospects for discovery of the pseudoscalar. 

\begin{figure}
	\includegraphics[keepaspectratio,width=0.5\textwidth]{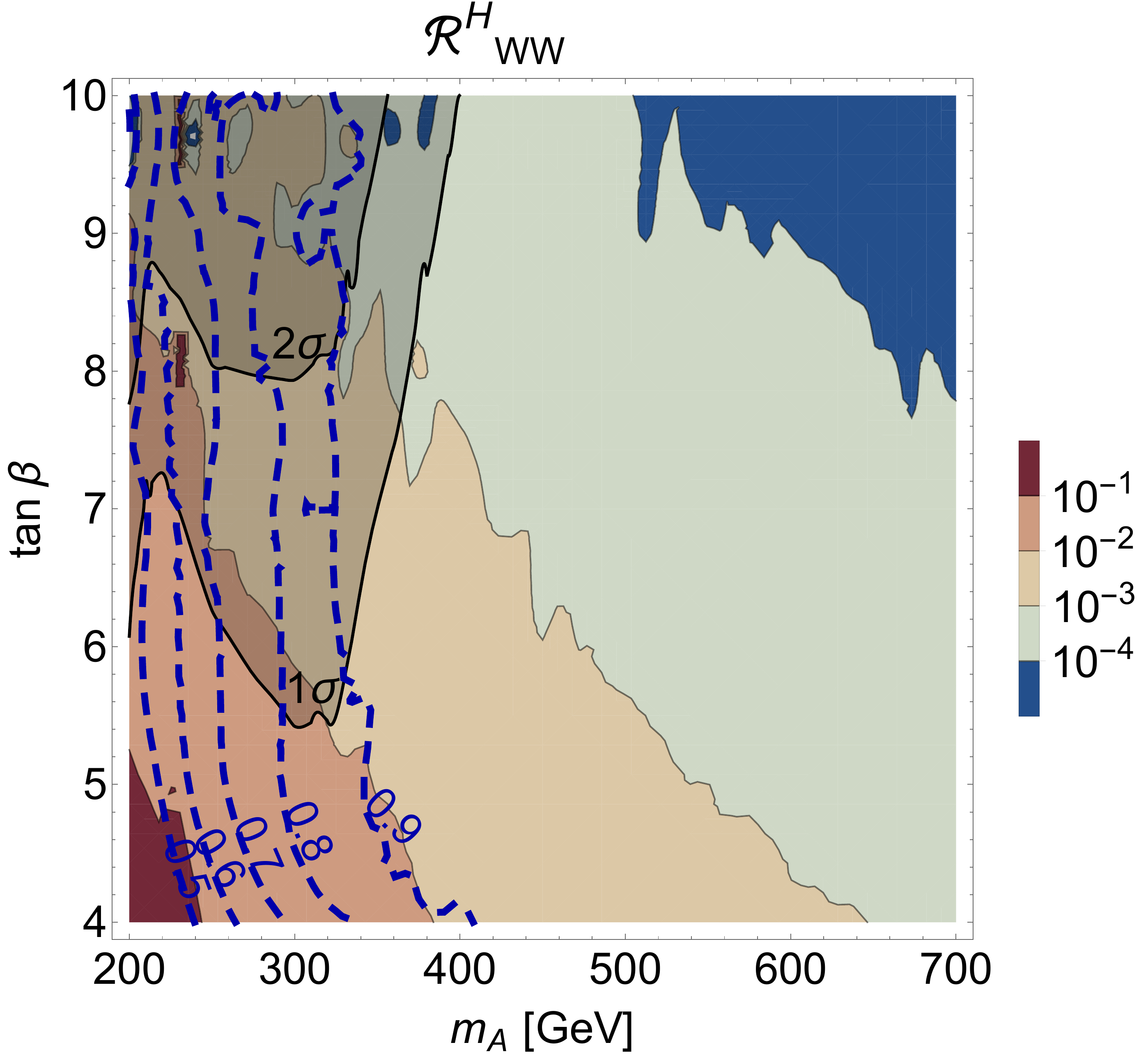}
	\caption{\label{HWWlimits}
		Shaded contours denote values of $\mathcal{R}^H_{WW}$. Gray shaded regions bounded by solid black lines show 1-$\sigma$ and 2-$\sigma$ exclusions by the $H/A \rightarrow \tau^+ \tau^-$ searches from the 8 TeV LHC run (excluded above). The dashed blue lines correspond to contours of $\mathcal{R}_{WW}^h$.
	}
\end{figure}

For the heavier CP-even Higgs $H$, in addition to the $\tau^+ \tau^-$ channel, there are non-negligible branching ratios into $WW$ or $hh$ despite being suppressed due to alignment (recall that, close to alignment, $H\approx H_{NSM}$). These branching ratios are largest at low $\tan\beta$ below the top mass threshold, whereas Br($H\rightarrow \tau^+ \tau^-$) is larger at higher $\tan\beta$. Note again that in the low $\tan\beta$ region, the main production of $H$ is via gluon fusion, which is enhanced due to the large unsuppressed top coupling. We computed the rate of $H\rightarrow WW$ relative to the SM expectation, $\mathcal{R}^H_{WW}$, which is shown as colored contours in Fig.~\ref{HWWlimits}. Current bounds on $\mathcal{R}^H_{WW}$ are at the level of $0.05-0.25$~\cite{Pelliccioni:2015hva}, hence dedicated searches at the LHC could probe the GCE best-fit regions, particularly for $m_A\lsim 350$ GeV, where $\mathcal{R}^H_{WW}$ can be within a factor of 10 of the current exclusion limit \cite{ATLAS-PHYS-PUB-2013-016,CMSPASFTR-13-024}. 

For $H/A$ heavier than about 350 GeV and low values of $\tan\beta$ ($\lsim 7$), both the CP-odd and even Higgs bosons preferentially decay to top quark pairs. However, due to the large SM $t\bar{t}$ background, this is a very challenging signature for the LHC~\cite{Craig:2015jba,Hajer:2015gka}; nevertheless, stronger sensitivity is expected at a 100 TeV collider~\cite{Hajer:2015gka}. The standard $\tau^+\tau^-$ searches can probe regions with larger values of $\tan\beta$. 

It should be kept in mind that, in addition to these searches for heavier Higgs bosons, the good fit regions at low $m_A\lsim 350$ GeV also predict deviations in $R^h_{WW}$ (see Eq.~\ref{RWW} for definition) at the 10\% level or more, hence such deviations from SM-like properties of the 125 GeV Higgs could be a stark signal of this scenario. All of the above search modes as well as the precision measurements of the 125 GeV Higgs are expected to improve substantially in sensitivity with the higher luminosity and energy of the 13 TeV LHC~\cite{ATL-PHYS-PUB-2014-016, CMS:2013xfa}.

\subsection{Direct Detection}
	
\begin{figure}
	\includegraphics[keepaspectratio,width=0.49\textwidth]{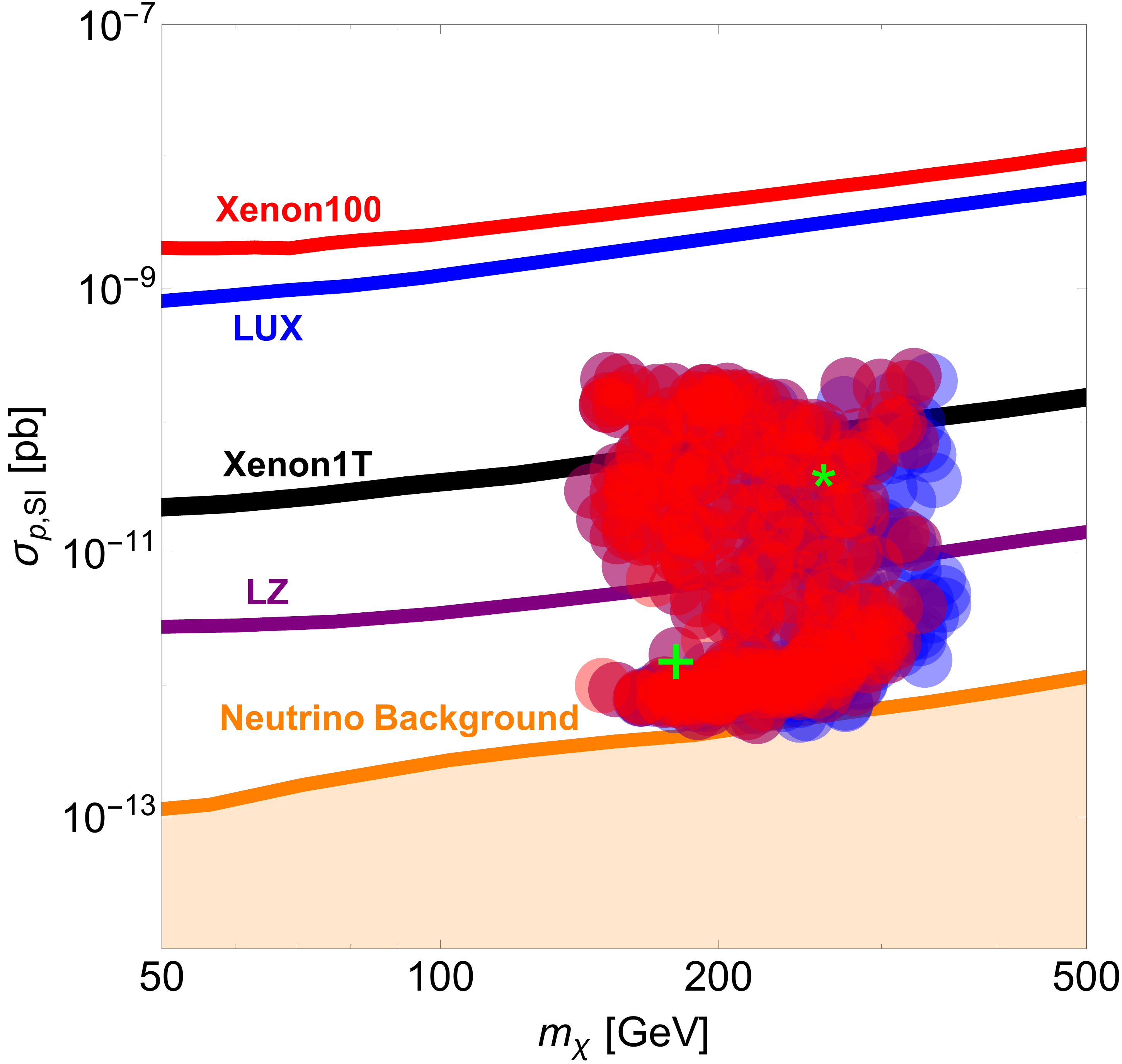}
	\caption{\label{fig:direct detection}
		Dark matter masses, $m_\chi$, and spin-independent DM-nucleon (proton) direct detection cross sections, $\sigma^p_{SI}$, predicted by our fits to the Fermi GCE. Points compatible with Fermi spectrum b (d) are in blue (red); we have only plotted points with $\chi^2\leq 50$ and compatible with collider and Higgs data (see text). The green cross and star correspond to the best fit points for spectrum (b) and (d) respectively.  Fig.~\ref{fig:chi2} shows $\chi^2$ contour regions from fitting the galactic center excess to Fermi spectrum (b) and (d). Current bounds (Xenon100, LUX), the reach of upcoming detectors (Xenon1T, LZ), and the neutrino background floor are also shown \cite{Feng:2014uja}.
	}
\end{figure}

Our predictions for spin-independent direct detection experiments are plotted in Fig.\,\ref{fig:direct detection}, which shows DM masses and spin-independent DM-nucleon (proton) direct detection cross sections compatible with the GCE (Fermi spectrum (b) in blue, spectrum (d) in red). We only show points with $\chi^2\leq 50$ that are compatible with both the $2\sigma$ $A/H \rightarrow \tau^+ \tau^-$ 8 TeV LHC constraints and $0.7 \leq \mathcal{R}_{WW}^h \leq 1.3$. As discussed in Section \ref{sec:directdetection}, we see that DM via the pseudoscalar resonance corresponds to generic cross sections of $\mathcal{O}(10^{-11})$pb, and these are comfortably safe from the existing Xenon100\,\cite{Aprile:2013doa} and LUX\,\cite{Akerib:2013tjd} bounds. A major fraction of the predicted parameter space can be probed with the next generation of direct detection experiments such as Xenon1T and LZ\,\cite{Feng:2014uja}. We note that almost all points predicted from our fit lie above the neutrino floor and therefore a signal can in principle be detected. The green cross and star correspond to the best fit points from Fig.\,\ref{fig:chi2} for spectrum (b) and (d) respectively. 

\section{Summary}
To conclude, we summarize the main findings of this paper:
\begin{itemize}

\item Recent reanalysis of GC background has found that the GCE could be consistent with annihilation of DM with much higher masses \cite{simonatalk, Agrawal:2014oha, Caron:2015wda, Bertone:2015tza}. This allows the GCE to be explained by the MSSM pseudoscalar resonance or ``$A$-funnel". We fit to two different dark matter spectra, Fermi spectrum (b) and (d) from \cite{simonatalk, Agrawal:2014oha}, and find that reasonable fits can be obtained while maintaining consistency with stringent constraints from collider searches, Higgs data, and direct and indirect detection.

\item For spectrum (b), the best fit region corresponds to $350$ GeV $\lsim m_A\lsim 450$ GeV and tan$\beta\lsim 6$. This region can be probed with searches for $H\rightarrow WW$ and $t\bar{t}$ resonance searches. $m_A\lsim 250$ GeV also gives reasonable fits but is incompatible with Higgs data.

\item For spectrum (d), there are two regions with reasonable fits to the GCE: $450$ GeV $\lsim m_A\lsim 600$ GeV at tan$\beta\lsim 8$, and $m_A\sim 300$ and tan$\beta\lsim 5.5$. The former region can yield signals at the LHC in the $A/H\rightarrow\tau\tau$ or $t\bar{t}$ resonance searches at the LHC. The latter region can also be probed with the same channels, and should also lead to measurements of deviations of the 125 GeV Higgs couplings from SM-like values. 

\item The best fit regions for both spectra (b) and (d) predict spin-independent direct detection cross sections of $\mathcal{O}(10^{-11})$pb for a $110$ GeV $\lsim m_\chi \lsim 350$ GeV neutralino. The entire region lies above the neutrino background, and the majority of the region is within reach of Xenon1T and LZ (see Fig.\,\ref{fig:direct detection}).
\end{itemize}

This exercise therefore leads to very sharp predictions for the next round of the LHC and direct detection experiments. Although the best fits obtained in this paper are noticeably worse than the best fit dark matter scenarios discussed elsewhere in literature, this highly predictive framework, coupled with the wide popularity of the MSSM, makes these results noteworthy. Even if the GCE turns out to be incompatible with the MSSM pseudoscalar resonance and is ultimately explained by some other (dark matter or astrophysical) phenomenon, this study still serves as a valuable template for the interplay between existing collider and Higgs constraints and the indirect, direct, and collider signatures of the $A$-funnel region with a light pseudoscalar in the MSSM. 

\medskip
\acknowledgements
It is our pleasure to thank Prateek Agrawal, Patrick Fox, Dan Hooper, Simona Murgia, Christopher Savage, and Carlos Wagner for valuable conversations. This work is supported by the U.S. Department of Energy, Office of Science, Office of High Energy Physics under Award Number DE-SC0007859. KF and AL are grateful for financial support from the Swedish Research Council (VR) through the Oskar Klein Centre and Stockholm University. AL is also supported by the Department of Physics at the University of Michigan, Ann Arbor. NS thanks the Aspen Center for Physics and the NSF Grant \#1066293 for hospitality during the completion of this work. NS is supported by a Wayne State University Research Grant Program award.

\newpage

\appendix

\section{Parameters and Vacuum Metastability}
\label{appendix1}

\begin{figure*}[h]	
	\includegraphics[keepaspectratio,width=0.49\textwidth]{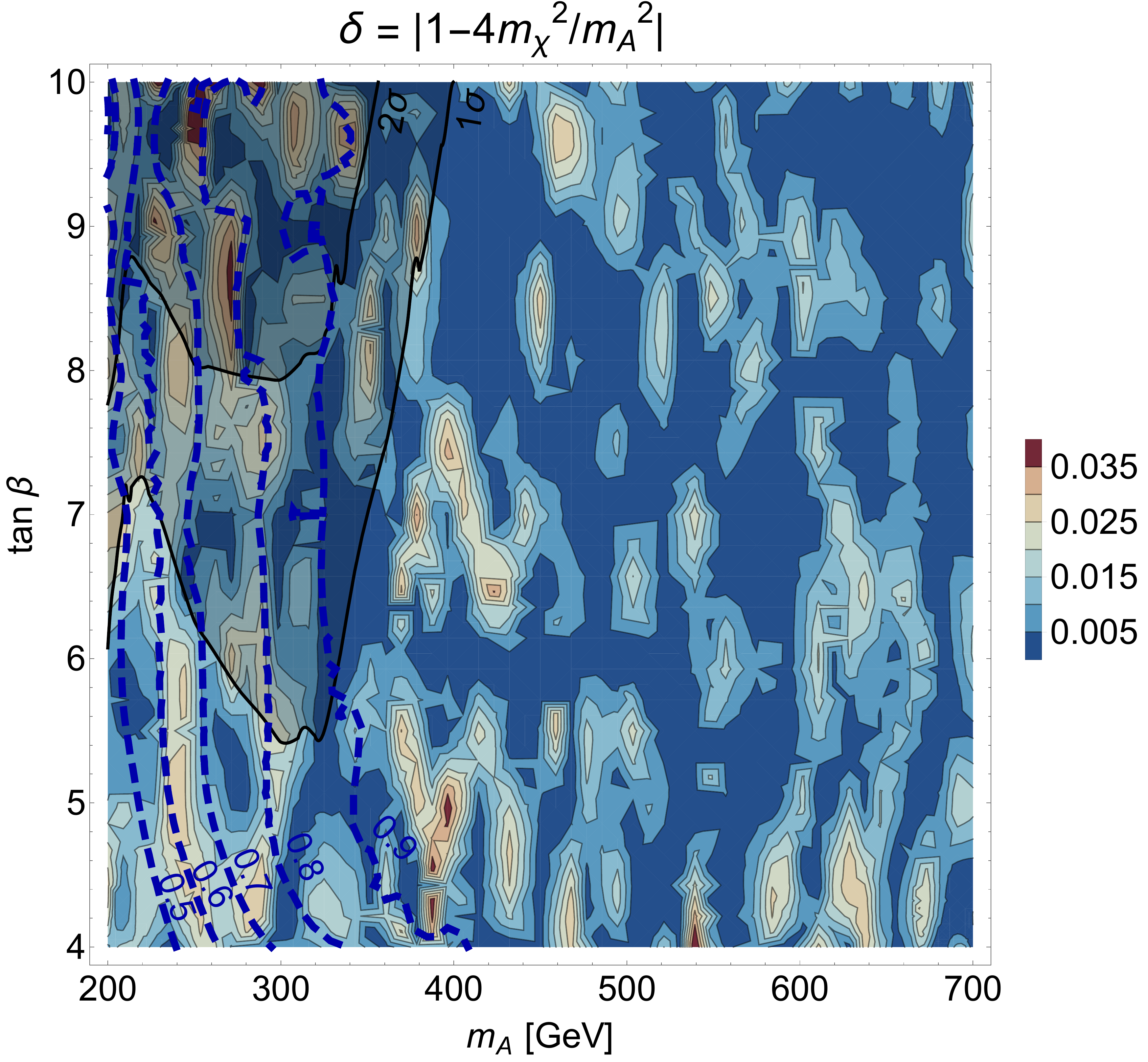}
	\includegraphics[keepaspectratio,width=0.45\textwidth]{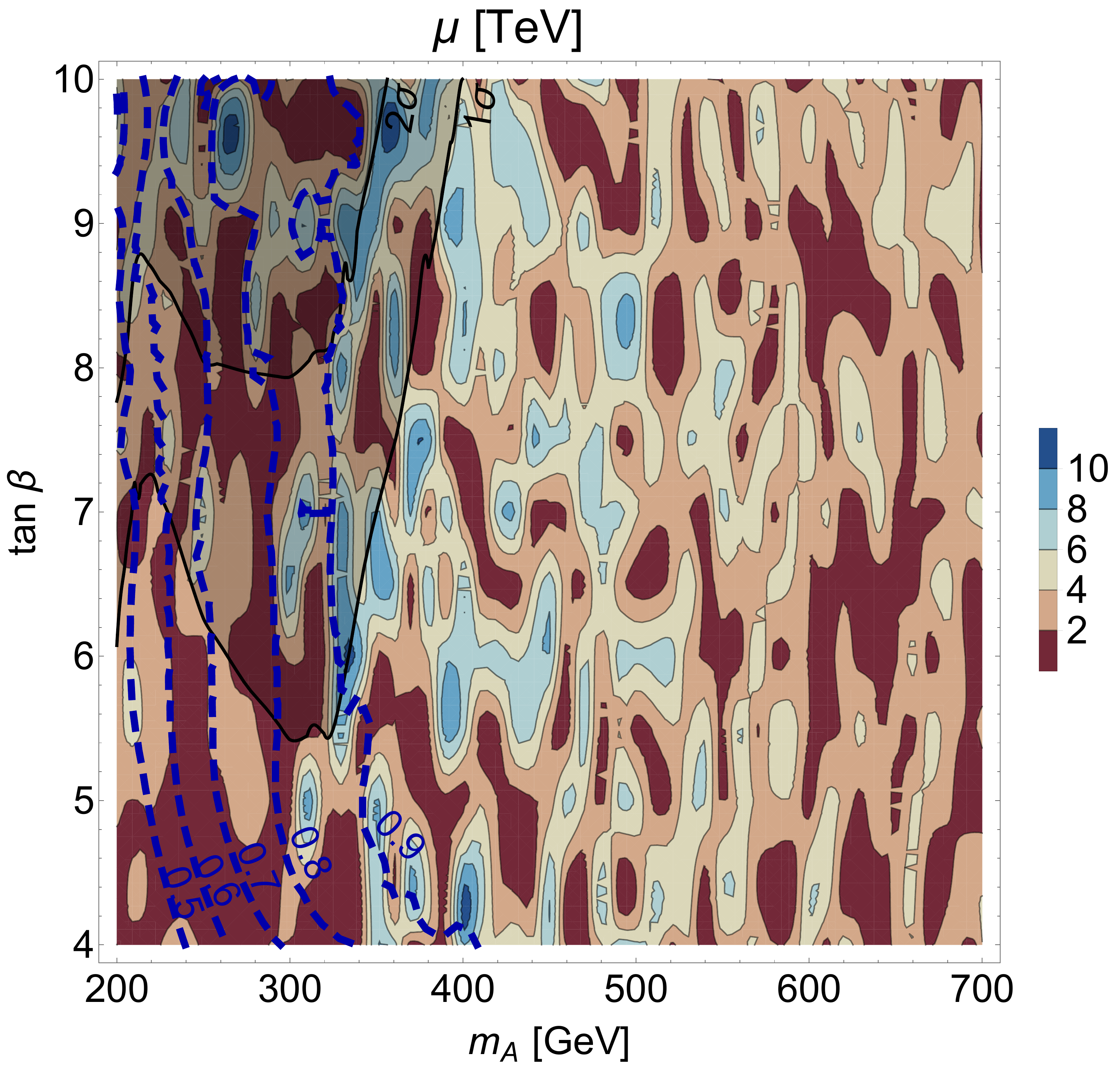}\\
	\vskip0.4cm
	\includegraphics[keepaspectratio,width=0.49\textwidth]{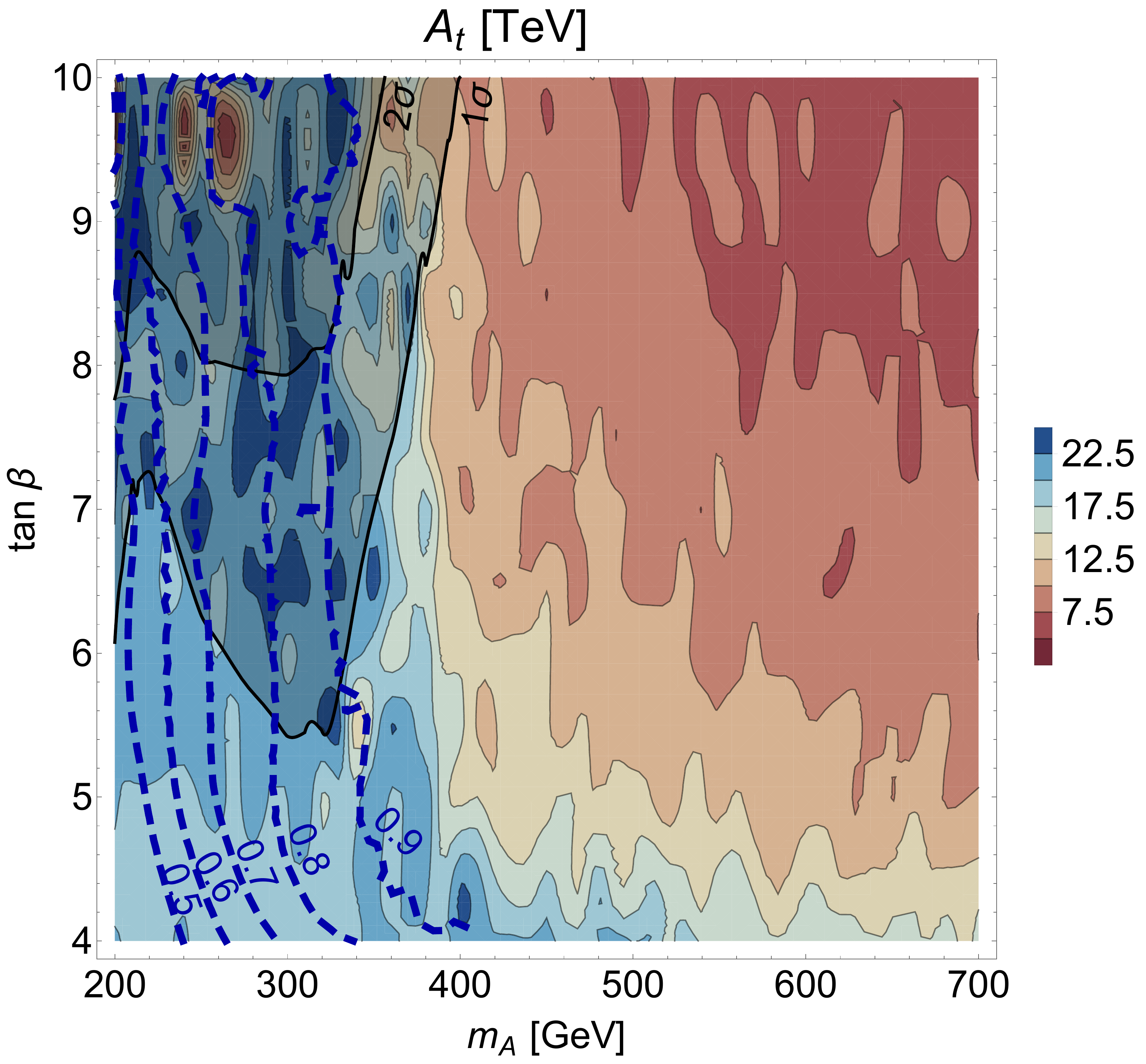}
	~~~~~\includegraphics[keepaspectratio,width=0.45\textwidth]{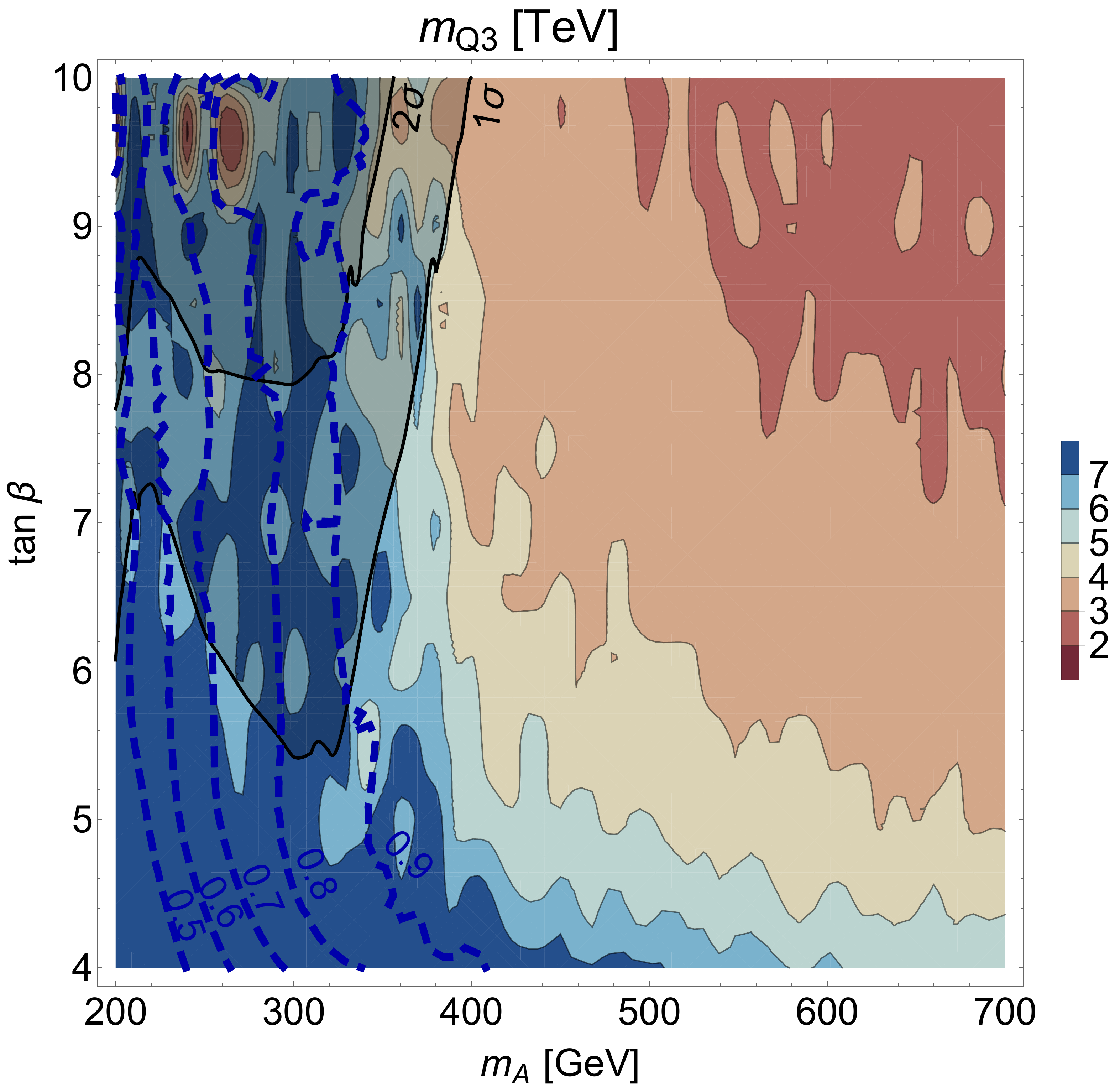}
	\hspace{\stretch{1}}	
	\caption{\label{fig:Param}
		Contours of various input parameter values in the scan region. See text for details. 
	}
\end{figure*}

\begin{figure}
	\includegraphics[keepaspectratio,width=0.49\textwidth]{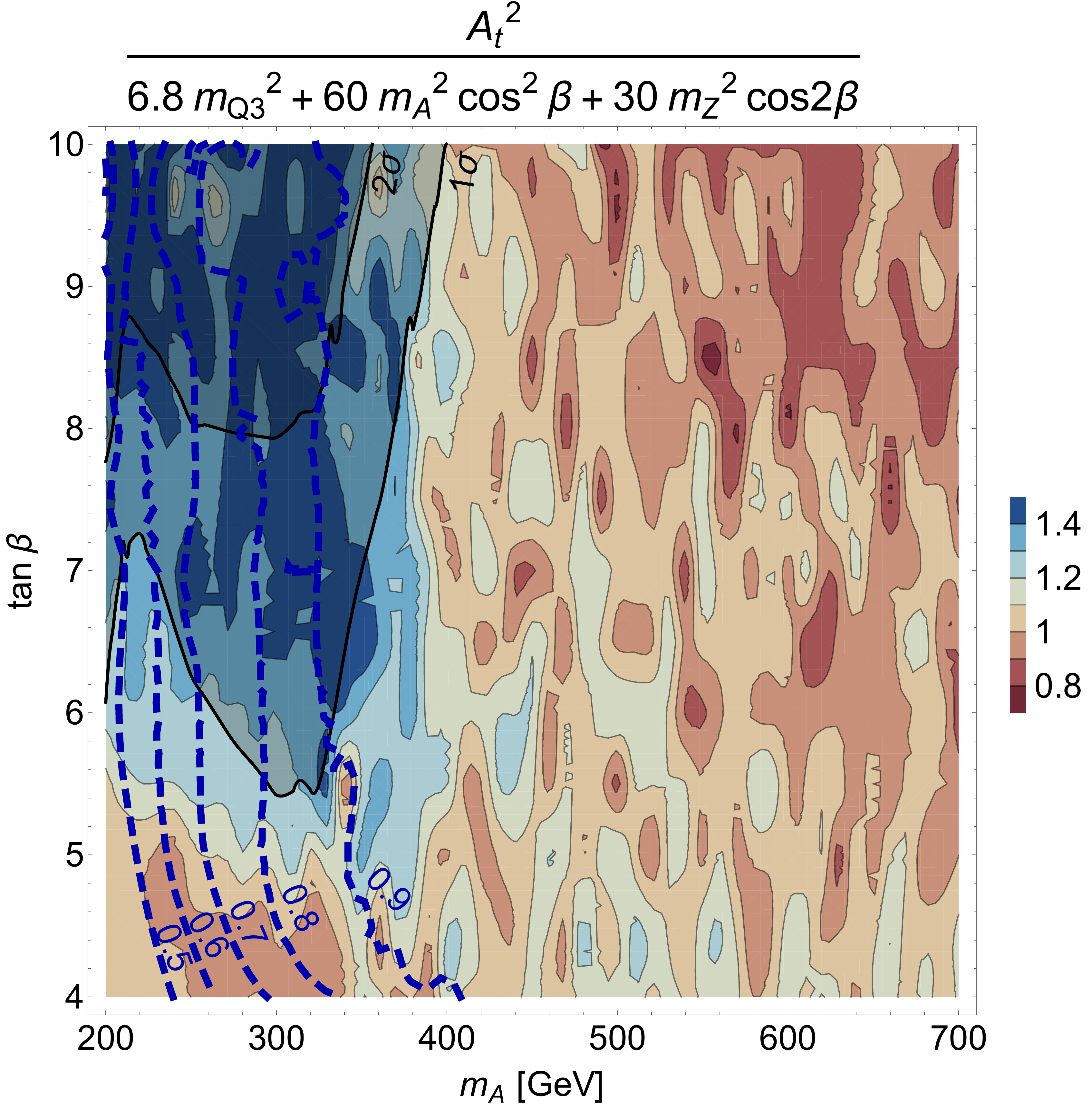}
	\caption{\label{atmqratio}
		Vacuum metastability requires this ratio to be approximately less than 1 \cite{Blinov:2013fta}, so we see that most of our points are compatible with vacuum metastability bounds.  
	}
\end{figure}

Fig.\,\ref{fig:Param} presents contour plots of the scanned parameters in the $m_A$-$\tan\beta$ plane. The approximate check for vacuum metastability from Eq.~\ref{metastability} is shown in Fig.~\ref{atmqratio}. It is seen that the desired condition is satisfied~(corresponding to the plotted ratio being less than 1) in most of the parameter space not ruled out by the 8 TeV LHC $A/H\rightarrow\tau^+ \tau^-$ bound. 

\newpage

\bibliography{gevresonance}

\end{document}